\documentclass[fleqn,usenatbib]{mnras}

\usepackage{newtxtext,newtxmath}
\usepackage[T1]{fontenc}

\DeclareRobustCommand{\VAN}[3]{#2}
\let\VANthebibliography\thebibliography
\def\thebibliography{\DeclareRobustCommand{\VAN}[3]{##3}\VANthebibliography}

\usepackage{graphicx}	% Including figure files
\usepackage{amsmath}	% Advanced maths commands
\usepackage{orcidlink}
\usepackage{bm}
\usepackage{makecell}
\usepackage{tikz}
\usepackage{booktabs}
\usetikzlibrary{shapes.geometric}
\usetikzlibrary{positioning, arrows.meta}  
\usetikzlibrary{shapes}    
\usetikzlibrary{shapes.multipart} 
\usetikzlibrary{decorations.pathreplacing} 
\usetikzlibrary{positioning, shapes.geometric}
\usetikzlibrary{calc} 
\usetikzlibrary{fit}

\newcommand{\meanKluge}{56.8}
\newcommand{\sigmameanKluge}{29.3}

\newcommand{\mumeanKluge}{24.5}
\newcommand{\sigmamumeanKluge}{1.3}
\newcommand{\meanhscgri}{74.4}
\newcommand{\sigmameanhscgri}{19.3}
\newcommand{\meanhscg}{70.5}
\newcommand{\sigmameanhscg}{32.8}
\newcommand{\meanhscr}{84.5}
\newcommand{\sigmameanhscr}{30.3}
\newcommand{\meanhsci}{69.3}
\newcommand{\sigmameanhsci}{27.6}

\newcommand{\mumeanhscg}{26.7}
\newcommand{\sigmamumeanhscg}{1.6}
\newcommand{\mumeanhscr}{26.2}
\newcommand{\sigmamumeanhscr}{1.5}
\newcommand{\mumeanhsci}{25.1}
\newcommand{\sigmamumeanhsci}{1.3}

\newcommand{\meanhydrangeag}{68.8}
\newcommand{\sigmameanhydrangeag}{28.9}
\newcommand{\meanhydrangear}{67.2}
\newcommand{\sigmameanhydrangear}{31.2}
\newcommand{\meanhydrangeai}{63.8}
\newcommand{\sigmameanhydrangeai}{24.2}
\newcommand{\meanhydrangeagri}{64.4}
\newcommand{\sigmameanhydrangeagri}{16.6}

\newcommand{\mumeanhydrangeag}{24.4}
\newcommand{\sigmamumeanhydrangeag}{1.3}
\newcommand{\mumeanhydrangear}{24.1}
\newcommand{\sigmamumeanhydrangear}{1.4}
\newcommand{\mumeanhydrangeai}{23.8}
\newcommand{\sigmamumeanhydrangeai}{1.4}

\newcommand{\meanhorizong}{59.8}
\newcommand{\sigmameanhorizong}{23.4}
\newcommand{\meanhorizonr}{65.3}
\newcommand{\sigmameanhorizonr}{39.5}
\newcommand{\meanhorizoni}{71.8}
\newcommand{\sigmameanhorizoni}{33.4}
\newcommand{\meanhorizongri}{69.0}
\newcommand{\sigmameanhorizongri}{27.3}

\newcommand{\mumeanhorizong}{24.0}
\newcommand{\sigmamumeanhorizong}{1.1}
\newcommand{\mumeanhorizonr}{23.1}
\newcommand{\sigmamumeanhorizonr}{1.3}
\newcommand{\mumeanhorizoni}{22.9}
\newcommand{\sigmamumeanhorizoni}{1.4}

\title[The BCG-ICL transition]{Isophotal signatures of the brightest cluster galaxy–intracluster light transition}

\author[Giulia Golini et al.]{
Giulia Golini \orcidlink{0009-0001-2377-272X}$^{1}$\thanks{E-mail: g.golini@unsw.edu.au},
Sarah Brough \orcidlink{0000-0002-9796-1363}$^{1}$,
Louisa Canepa\orcidlink{0009-0002-5450-6683}$^{1}$,
Garreth Martin\orcidlink{0000-0003-2939-8668}$^{2}$,
and Matthias Kluge\orcidlink{0000-0002-9618-2552}$^{3}$
\\
$^{1}$School of Physics, The University of New South Wales (UNSW Sydney), Sydney, NSW 2052, Australia\\
$^{2}$School of Physics $\&$ Astronomy, University of Nottingham, University Park, Nottingham NG7 2RD, UK\\
$^{3}$Max-Planck Institute for Extraterrestrial Physics, Giessenbachstrasse 1, D-85748, Garching, Germany
}

\date{Accepted 2026 October 8. Received 2026 September 21; in original form 2026 July 22
}

\pubyear{\the\year{}}

\begin{document}
\label{firstpage}
\pagerange{\pageref{firstpage}--\pageref{lastpage}}
\maketitle

\begin{abstract}

The distinction between brightest cluster galaxies (BCGs) and intracluster light (ICL) remains ambiguous, with inferred cluster properties depending strongly on the adopted definition and measurement methodology.
We report the presence of a common inflection feature in the radial profiles of surface brightness, ellipticity, and position angle of BCG+ICL systems, indicative of a structural transition in the underlying stellar distribution. Using a combination of low-redshift clusters ($z < 0.1$) observed with the Wendelstein Wide Field Imager, deep Hyper Suprime-Cam observations of clusters at $0.2 < z < 0.4$, and cosmological hydrodynamical simulations from Horizon-AGN and Hydrangea, we consistently identify a common inflection at radii of  $60-85$ kpc. We interpret this feature as a characteristic transition in the combined BCG+ICL light distribution, that may reflect the hierarchical assembly of the cluster.
The persistence of this feature across multiple photometric bands, redshift ranges, and simulations supports its robustness. To identify these transitions, we employ an automated second-derivative-based approach that does not rely on parametric decompositions or arbitrary surface brightness thresholds.
This framework provides a physically motivated and scalable means to characterise the BCG–ICL transition, and it will be of paramount importance for forthcoming large-scale surveys such as the Legacy Survey of Space and Time and the European Space Agency’s Euclid mission.

\end{abstract}

\begin{keywords}
galaxies: clusters: general -- galaxies: evolution -- galaxies: photometry

\end{keywords}

%%%%%%%%%%%%%%%%%%%%%%%%%%%%%%%%%%%%%%%%%%%%%%%%%%

%%%%%%%%%%%%%%%%% BODY OF PAPER %%%%%%%%%%%%%%%%%%

\section{Introduction}

Galaxies in clusters are embedded within a common diffuse luminous envelope \citep{Zwicky1937, zwicky1951coma,zwicky1952luminous,zwicky1957non}, now referred to as intracluster light \citep[ICL; see][for reviews]{mihos2015intragroup, contini2021origin,montes2022faint,2024Contini}. This envelope is composed of stars that are not gravitationally bound to individual satellite galaxies, but are rather associated to the Brightest Cluster Galaxy (BCG) and to cluster gravitational potential \citep{Dressler1979, Arnaboldi1996, Longobardi2018}.

The formation of this diffuse stellar component is primarily driven by hierarchical structure assembly. Both minor and major mergers involving central and satellite galaxies \citep{monaco2006diffuse, Murante, gerhard2007kinematics, Contini2014, jimenez2018unveiling, jimenez2019j, groenewald2017close,joo2023intracluster}, as well as tidal stripping processes \citep{Rudick2009, Rudick2011, Contini2014, demaio2015origin, monte2018, jimenez2018unveiling, jimenez2019j}, are widely regarded as the dominant channels contributing to its build-up. Pre-processing within infalling galaxy groups can also play a significant role \citep{willman2004origin, mihos2005diffuse, rudick2006formation}. Other mechanisms, such as in-situ star formation \citep{puchwein2010intracluster} and dwarf galaxy disruption \citep{Murante, Purcell2007, Conroy2007, Martel2012, giallongo2014diffuse, Annunziatella2016, Morishita2017, raj2020fornax}, are thought to contribute only to a small degree \citep[e.g.,][]{melnick2012intergalactic, Contini2014, demaio2015origin, monte2018}.
Isophotal analyses of the ICL have further revealed radial trends in ellipticity \citep[see e.g.,][]{Gonzalez2005,kluge2020structure, montes2021buildup} and a close alignment with the cluster large-scale structure and dark matter (DM) distribution \citep[e.g.,][]{Despali2014, Montes2019, alonso2020intracluster, deason2021stellar, lau2021correlations, montes2022new, diego2023exploring, contreras2024characterising,yoo2024spatial, asensio2025intracluster,butler2025intracluster, ellien2025euclid,fernandez2026intracluster}. 
These findings point to a complex component that encodes key information about the formation and assembly history of galaxy clusters.

Deep imaging observations have shown that the BCG and the ICL form a continuous, extended stellar component that can be traced out to several hundreds kiloparsecs down to a surface brightness of $\sim 30$–32 mag/arcsec$^{2}$ \citep{montes2014intracluster,zhang2019dark,kluge2020structure}. This continuity implies a gradual rather than sharp transition between the BCG and ICL, which are not separated by a distinct physical boundary but instead likely represent spatially overlapping stellar populations with different assembly histories (Pearce et al. (in preparation)), making their separation challenging.
Despite the efforts to characterize the evolution of the BCG and ICL as separate components, their decomposition remains a persistent problem in astrophysics because it is fundamentally dependent on their definition.
From an observational perspective, various techniques have been used to distinguish between the BCG and ICL.
These include surface brightness thresholds \citep[e.g.,][]{feldmeier2004deep, Presotto2014,montes2014intracluster, burke2015coevolution,monte2018,montes2021buildup,furnell2021growth,mayes2026coevolution}, non-parametric approaches \citep[e.g.,][]{gonzalez2007census, demaio2018lost, martinez2023galaxy,butler2025intracluster}, composite-model decompositions \citep[e.g.,][]{Gonzalez2005,Presotto2014,Iodice2016,spavone2017vegas, monte2018,cattapan2019vegas,iodice2020vegas, spavone2020fornax, raj2020fornax, kluge2021photometric,montes2021buildup, ragusa2021vegas,ragusa2022intra, kluge2023minor,martinez2023galaxy,joo2023intracluster, Ahad2023, ragusa2023does}, multi-galaxy fitting \citep[e.g.,][]{giallongo2014diffuse, jimenez2016disentangling, Morishita2017,jimenez2018unveiling,poliakov2021quantified}, wavelet-based decompositions \citep[e.g.,][]{da2005intragroup,Adami2012, guennou2012intracluster,ellien2021dawis}, kinematic studies of planetary nebulae and globular cluster populations \citep[e.g.,][]{Arnaboldi1996, powalka2018next, Harris2020, Hartke2022, Kluge2023}, stacked integral-field spectroscopy \citep{Edwards2016, Edwards2020} and image stacking \citep[e.g.,][]{Zibetti2005, zhang2019dark, SampaioSantos2021, chen2022sphere, golden2023characterizing}.
Similar techniques have also been adopted in cosmological simulations: aperture-based criteria \citep{pillepich2018simulating}, kinematic decompositions \citep{Dolag2010, Remus2017, Proctor2024}, binding-energy methods \citep{willman2004origin, Murante,Dolag2010}, density thresholds \citep{Rudick2009}, and again, surface brightness-based definitions \citep{feldmeier2004deep,mihos2005diffuse,Rudick2010}.

The various approaches outlined above correspond to different definitions of the two components. These differing definitions result in distinct decompositions and, consequently, to different inferred ICL properties for the same cluster \citep{brough2024preparing}.
As a result, the attention has shifted away from establishing a distinct BCG–ICL boundary or separating the two components, to characterizing a transition region, where the ICL component begins to dominate over the stellar distribution associated with the BCG.
Observational estimates place this transition between $\sim$ 40 kpc to 80 kpc \citep{Zibetti2005, Iodice2016,gonzalez2021discovery, yoo2021intracluster, spavone2024galaxy, zhang2025splashback} up to 200 kpc \citep{chen2022sphere}.  Simulations predict a broad distribution spanning $\sim$ 15 to more than 100 kpc \citep{2018Pillephic, Proctor2024}, depending on the BCG morphology and on cluster dynamical state \citep{contini2021brightest, Contini2022, Contini2024}. 
%--------------
\cite{gonzalez2021discovery} defined the transition between the BCG-dominated to ICL-dominated region of the massive cluster MACS J1149.5$+$2223, by calculating the logarithmic derivative of its surface brightness profile. They identified an inflection in this gradient at $\sim$ 70 kpc.
Similarly, \cite{zhang2025splashback} stacked over 4,000 dark energy survey (DES) galaxy clusters to obtain high signal-to-noise (SN) measurements of their average luminosity profile slope, defining the BCG-ICL transition as a dip in the radial logarithmic slope of the stacked surface brightness profiles. For the full sample, the dip minimum occurred at 40-60 kpc and shifted slightly outward with increasing cluster richness. They interpreted this feature as the physical edge of the central galaxy and the onset of the diffuse, actively accreting light component, identifying it as the galaxy's "splashback zone" where stellar particles reach their outermost apocenters before migrating inward.
%-------------------------------
\cite{gonzalez2021discovery} and \cite{zhang2025splashback} estimates, however, rely only on inflections of the surface brightness profiles of the BCG+ICL system.
Some studies (e.g., \citealp{montes2021buildup,marrero2026large}) indicate that the transition may coincide with not only a flattening of the surface brightness profile, but also with a radial increase in the isophotal ellipticity ($\epsilon$\footnote{Throughout the text, the ellipticity is defined as $\epsilon$ = 1 - a/b, where a and b are the semi-minor and semi-major axes of the ellipse, respectively.}) and position angle (PA) twists of the BCG+ICL system. These findings imply that the BCG–ICL transition could be seen in both the photometric and geometric structures of the system.
Yet, in the absence of systematic studies across large samples, it remains unclear whether these geometric variations are ubiquitous or whether they can serve as reproducible and observationally reliable tracers of transition in  BCG+ICL systems.
Addressing this question is particularly timely in the context of forthcoming wide-field surveys such as \textit{Euclid} \citep{laureijs2011euclid, scaramella2022euclid,bellhouse2025euclid} and the Vera C. Rubin Observatory Legacy Survey of Space and Time (LSST; \citealp{ivezic2019lsst}), which will deliver deep images for thousands of galaxy clusters. Therefore, automated methods capable of identifying the BCG–ICL transition across large samples are crucial for fully exploiting the scientific potential of these datasets.

%%%%%%%%%%%

In this study, we use isophotal fitting to derive the radial surface brightness, ellipticity and position angle profiles for different clusters. We combine cosmological hydrodynamic simulations with deep imaging to identify structural changes in the BCG+ICL system. The aim of the analysis of profiles derived from independent observational datasets and reduction pipelines is the assessment of whether the inferred inflection-point properties are intrinsic to the BCG+ICL systems rather than artefacts of a specific methodology.

This paper is structured as follows. In Sect.~\ref{sect:datasets}, we describe the simulation and observational datasets. In Sect.~\ref{sec:Methods}, we outline our methodology to locate the BCG-ICL transition. In Sect.~\ref{results}, we present our results on the identification of transition radii and check for correlations with host cluster parameters. In Sect.~\ref{sec:discussion}, we discuss the implications for the structure and assembly of the BCG+ICL system and discuss possible future approaches. Finally, in Sect.~\ref{conclusions}, we summarise our main conclusions.
Throughout this work, we adopt a flat lambda cold dark matter ($\Lambda$CDM) cosmology with H$_0 = 70\,\mathrm{km\,s^{-1}\,Mpc^{-1}}$ and matter density parameter $\Omega_{\rm m}=0.3$.

\section{Data and Image Preparation}
\label{sect:datasets}

\begin{figure}
  \centering
  \includegraphics[width=\linewidth]{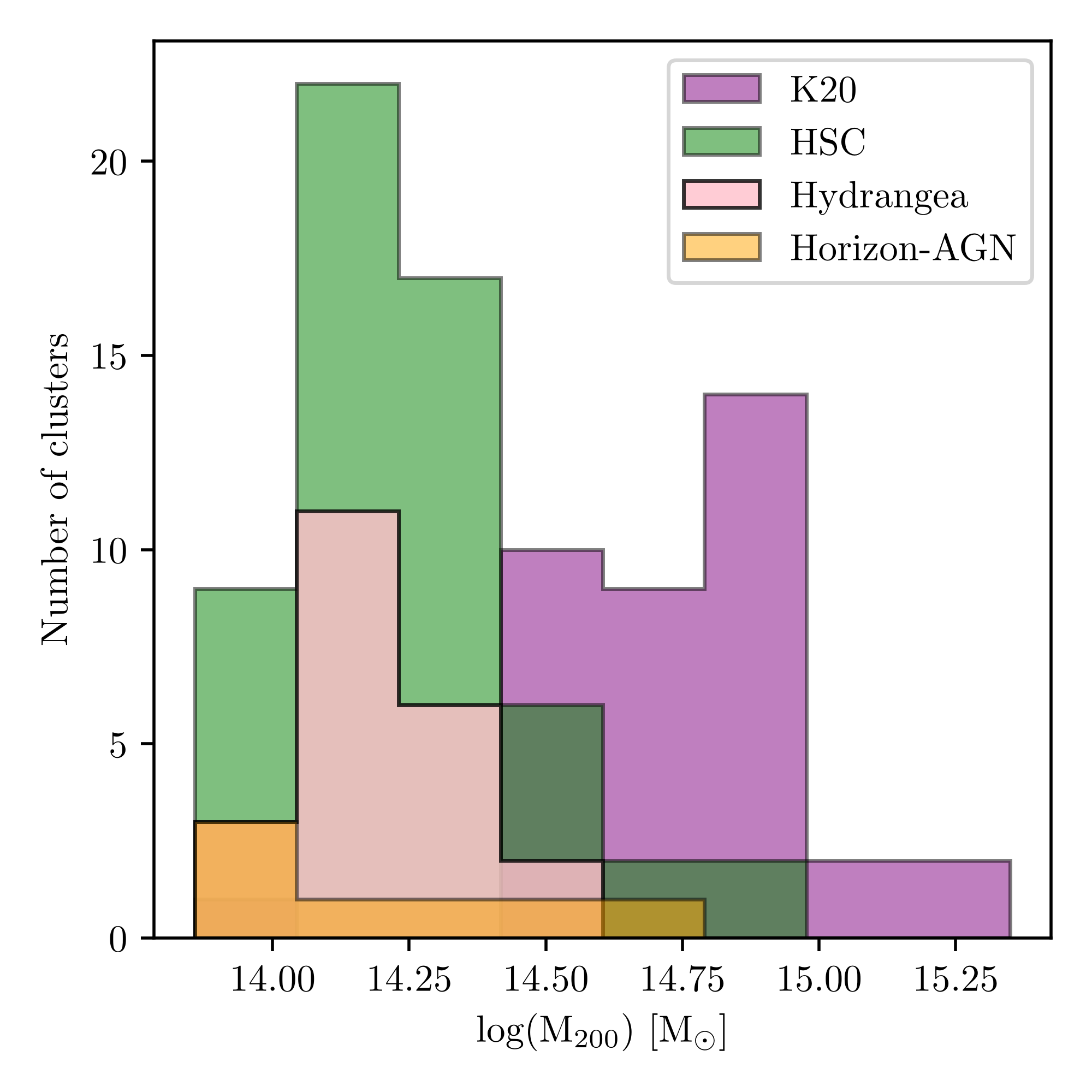}
  \caption{Distribution of the cluster masses for the four samples used in this work.}
  \label{fig:masses}
\end{figure}

In this work, we analyse four independent datasets comprising both simulated and observational samples of galaxy clusters. These datasets have been selected to resemble the 10-year LSST surface brightness limits of 30.3 mag/arcsec$^{2}$ (3$\sigma$; 10$\arcsec\times 10\arcsec$)\footnote{\href{https://smtn-016.lsst.io/}{https://smtn-016.lsst.io/}}. On the theoretical side, we consider the Horizon-AGN simulation \citep{Dubois2014} and the Hydrangea suite of cosmological zoom-in simulations \citep{Bahe2017}, which are introduced in more detail below but differ in terms of numerical implementation, resolution, box size, and subgrid prescriptions for galaxy formation and feedback.
On the observational side, we use ICL profile measurements of \citet{kluge2020structure}, hereafter K20, and deep imaging from the Hyper Suprime-Cam (HSC) survey. These datasets are characterised by different observation strategies and data reduction pipelines. Unlike the simulations, they are subject to systematic observational effects such as the point spread function (PSF), sky subtraction methods, data noise, and the need to mask other sources. All of these factors can impact the detection and characterisation of the intracluster light.  
By combining datasets spanning a wide range of physical assumptions, observational conditions, and analysis techniques, our aim is to test the robustness of the identified features and assess whether they persist independently of specific methodologies or systematic biases.
The mass distribution of the clusters used in this work is shown in Fig.~\ref{fig:masses} and the main properties of the four datasets are summarised in the following section.

\subsection{Cosmological Simulations}
\subsubsection{Hydrangea}
\label{subsubsec:hydrangea}
Hydrangea\footnote{\href{https://virgo.dur.ac.uk/2017/06/14/Hydrangea/}{https://virgo.dur.ac.uk/2017/06/14/Hydrangea/}} \citep{Bahe2017} is a suite of cosmological hydrodynamical zoom‑in simulations of massive galaxy clusters, based on a variant of the Evolution and Assembly of GaLaxies and their Environments (EAGLE) galaxy formation and evolution model \citep{Schaye2015, Bahe2017, bahe2019disruption}. The simulations use the Smoothed Particle Hydrodynamics (SPH) code  GADGET‑3 \citep{Springel2005} and include sub‑grid prescriptions for gas cooling, star formation, stellar and active galactic nuclei (AGN) feedback, calibrated to reproduce observed stellar masses, sizes, and gas fractions of group‑scale haloes. The simulations reproduce the satellite stellar mass function well from the local Universe to z $\sim$ 1.5, and the total stellar mass of clusters is realistic, although the BCGs are known to be over‑massive by a factor of 2 to 3 \citep{Bahe2017}. The enhanced substructure identification scheme removes stars bound to satellites more rigorously than the standard Subfind method, resulting in lower inferred BCG and ICL stellar masses.

\subsubsection{Horizon-AGN}
\label{subsubsec:horizonagn}

Horizon‑AGN\footnote{\href{https://www.horizon-simulation.org/}{https://www.horizon-simulation.org/}} \citep{Dubois2014,dubois2016horizon,kaviraj2017horizon} is a cosmological‑volume hydrodynamical simulation performed with the adaptive mesh refinement (AMR) code \textsc{RAMSES} \citep{Teyssier2002}. The simulation follows a cubic volume with a side length of 142 Mpc, initially sampled by 1024$^{3}$ DM particles with a mass resolution of 8 $\times 10^{7}\,\rm M_\odot$. The gas is initially discretised on a uniform grid of size 1024$^{3}$ and refined according to a quasi‑Lagrangian criterion down to a minimum physical cell size of 1 kpc. Sub‑grid physics includes radiative cooling from hydrogen, helium, and metals down to 10$^4$K \citep{1993Sutherland}, star formation following a Schmidt law \citep{Kennicutt1998}, and feedback from Type Ia and II supernovae and stellar winds. Stellar particles, which constitute the ICL component, have a mass resolution of $\sim 2 \times 10^{6}\,\rm M_\odot$. We select clusters with 13.9 M$_{\odot}$< log(M$_{200}$) < 14.8 M$_{\odot}$.
For this study, we exclude clusters with multiple BCGs because their overlapping light distributions make the BCG-ICL transition more difficult to identify. To validate our method, we focus on simpler systems with a single dominant BCG.

\subsubsection{Mock images}
\label{subsec:Mockimages}

In order to work with simulated data, we need to create synthetic mock observations. This is one of the most robust ways to compare simulations with observational data \citep[e.g.][]{Jonsson2006, Naab2014, Choi_2018, CAMPS2020100381}. 
To make these synthetic observations we use the method described in \cite{martin2022preparing} and summarised in what follows. 

We generate idealized mock images for each simulated cluster by extracting all star particles within a $(4 \mathrm{Mpc})^3$ region centred on the BCG. The spectral energy distribution (SED) of each particle is computed using the \cite{Bruzual2003} stellar population models assuming a \cite{Chabrier2003} initial mass function (IMF) based on the particle age and metallicity.
Dust attenuation is neglected, as the diffuse ICL is expected to contain little dust and dust modelling would depend strongly on the different feedback and hydrodynamical prescriptions adopted by each simulation. Particle luminosities are redshifted to $z=0.1$ and convolved with the HSC filter transmission curves. To better represent the stellar mass distribution and reduce pixel-scale noise, we apply an adaptive smoothing scheme similar to the adaptiveBox method of \cite{Merritt2020}.

The final mock images are constructed by projecting stellar particles onto a two-dimensional grid with a pixel scale of 0.168$''$/pix $\times$ 0.168$''$/pix. This choice is motivated by the need to facilitate direct comparison with HSC data, that has the same pixel scale and to have a comparable resolution of the data that the LSST camera will deliver (0.2$''$/pix).
For each cluster, images are generated in three projections, convolved with the HSC PSF, and Gaussian noise is added to reproduce the LSST 10 year limiting surface brightness of 30.3 mag/arcsec$^{2}$ (3$\sigma$; 10$\arcsec \times 10\arcsec$) estimated following the Rubin Observatory documentation available online\footnote{\href{https://smtn-016.lsst.io/}{https://smtn-016.lsst.io/}}. We use the $g$ band HSCam 2D PSF measured by \cite{montes2021buildup} out to 289$\arcsec$. This radial extent is enough to cover the inner 300 kpc of the simulated clusters at $z = 0.1$.
Since instrumental and astrophysical contaminants are neglected, these mock observations are representative of an idealized scenario for testing the BCG-ICL transition. 
%We show an example of a mock image from the Hydrangea simulation in the upper left panel of Fig.~\ref{fig:ex}.

\subsection{Observational sample}
\subsubsection{Wendelstein Wide Field Imager}
\label{subsubsec:klugedata}

We adopt deep $g$ band imaging of 170 nearby galaxy clusters from K20 \citep{kluge2020structure}, obtained using the Wendelstein Wide Field Imager (WWFI) on the 2m Fraunhofer telescope at the Wendelstein Observatory \citep{Kosyra2014}. The WWFI provides a field of view (FoV) of $27.6' \times 28.9'$ with a pixel scale of 0.2$\arcsec$/pix, allowing accurate measurements of both the extended ICL and the surrounding sky background. A large dithering strategy was adopted to minimise systematic effects and improve background subtraction. The sample is primarily drawn from the Abell–Corwin–Olowin (ACO; \citealp{Abell1989}) cluster catalogue and was selected to target massive nearby systems with typical halo masses $> 10^{14} \rm M_{\odot}$ estimated by assuming a linear relationship between gravitational mass (M$_g$) and M$_{200}$. The clusters span a redshift range of $0.02 < z < 0.14$ with a median redshift of $z=0.06$.
The surface brightness profiles of the 170 clusters are corrected for dust extinction using the maps
from \cite{schlafly2011measuring}, K-corrected using \cite{chilingarian2010analytical} and \cite{chilingarian2012universal}, and corrected for cosmological (1 + z)$^4$ dimming.
Moreover, PSF effects are mitigated by modeling the natural power-law decline of stellar wings to subtract scattered light from foreground stars and by employing a double-convolution technique to remove the light redistributed from the BCG’s central region into its outskirts.
The data reach a limiting surface brightness of $\mu_{g} \sim$ 30 mag/arcsec$^{2}$ limited by residual background uncertainties in the surface brightness profiles. Further details on the observation and data reduction can be found in \cite{kluge2020structure}.

\subsubsection{Hyper Suprime Camera}
\label{subsubsec:hsc}

The second observational dataset used in this work is taken from Hyper Suprime-Cam Subaru Strategic Program (HSC-SSP; \citealt{Aihara2018, Miyazaki2018}) Public Data Release 2 \citep[PDR2;][]{10.1093/pasj/psz103}. HSC is a wide-field imaging instrument on
the 8.2m Subaru telescope with a FoV of $1.77~\mathrm{deg}^2$, operated by the National Astronomical Observatory of Japan on Maunakea, Hawaii. The camera consists of 116 CCDs (104 science detectors, four guide sensors, and eight focus sensors) with a pixel scale of 0.168$\arcsec$/pix.

HSC-SSP is a layered multi-band ($grizy$) imaging survey consisting in three layers with varying limiting depths: Wide (m$_r \sim$ 26.4 mag), Deep
(m$_r \sim$ 27.4 mag), and Ultradeep (m$_r \sim$ 28.0 mag).
For this analysis, we use Ultradeep imaging data in the $g$, $r$, and $i$ bands, retrieved from the Subaru Mitaka Okayama Kiso Archive (SMOKA; \citealt{2002Baba}).

We use  the Cluster finding algorithm based on Multi-band Identification of Red-sequence gAlaxies (CAMIRA; \citealp{oguri2018optically}), an optically selected cluster catalogue constructed from the full HSC-SSP dataset. The algorithm used to produce the catalogue \citep{oguri2014cluster} exploits the presence of the "red sequence" galaxies in massive clusters \citep[e.g.,][]{gladders2000new} to identify likely member galaxies using a stellar population synthesis model and subsequently group them into clusters. The final CAMIRA catalogue provides cluster positions, with the cluster centre defined as the coordinates of the BCG identified by the algorithm, photometric redshifts, richness and stellar masses for each member galaxy. For this work, we restrict our analysis to clusters in the UltraDeep layer, whose depth is well suited to low-surface brightness studies \citep[$\mu_r \sim 29.8$ mag/arcsec$^{2}$;][]{martinez2023galaxy}.
Photometric redshifts are estimated by the CAMIRA algorithm from the most probable red sequence associated with each cluster \citep{oguri2014cluster}. We restrict our sample to clusters with $z > 0.2$ due to the large angular sizes and brightness of
member galaxies in HSC-SSP below this redshift complicating the cluster finding \citep{oguri2018optically} and $z < 0.4$, as the detection of low surface brightness features becomes increasingly difficult due to cosmological dimming.
In CAMIRA, richness is defined as the number of red-sequence member galaxies with stellar masses M$_\star \gtrsim 10^{10.2}\,\rm M_\odot$ within a projected radius of $\sim 1\,h^{-1}\,\mathrm{Mpc}$, where stellar masses are estimated using a stellar population synthesis model \citep{oguri2014cluster}.
Halo masses are derived from the CAMIRA richness values using the richness--mass scaling relation presented by \citet{murata2019mass}. 
We also remove clusters from these layers that have significant contamination from bright stars (identified by manual inspection of these images), or clusters with double BCGs.
The final sample consists of 60 clusters in the Ultradeep layer, in the redshift range $0.2 < z < 0.4$ and our final sample covers a halo mass range of $\sim10^{13.8}\,\rm M_\odot$ to $10^{15}\,\rm M_\odot$.

\subsection{Profile extraction}
\label{subsec:isofitting}

Before deriving the averaged surface brightness profiles and structural properties of the BCG+ICL systems through isophotal fitting, it is key to address contamination from foreground and background sources, as well as cluster members (masking) and local sky subtraction. In the following we describe how this is performed for the different samples.

\subsubsection{Masking and sky subtraction}
\label{subsec:skysub}

K20 employed an iterative approach to improve source detection: A preliminary model of the BCG+ICL system was first constructed and subtracted from the image, enabling the identification of small and intermediate-sized sources. Then, the resulting masks were expanded to account for extended emission, while bright or morphologically complex objects were masked manually. A constant sky level was then estimated and subtracted from the image \citep[see][for further details]{kluge2020structure}.

For HSC data, we first mask contaminating sources with a custom procedure. We use \texttt{Photutils} \citep{Bradley2019} to generate and dilate segmentation maps, we identify the central object (the BCG), and mask all the remaining detected sources while preserving the surrounding diffuse emission. The masking procedure is further refined using a custom faint-source catalogue and manual intervention using the software \texttt{Noisechisel} \citep{gnuastro, noisechisel_segment_2019}. 
Following source masking, we perform sky subtraction adopting the procedure described in Canepa et al. (in prep.). We fit elliptical isophotes using the \texttt{Photutils} implementation of \citet{Jedrzejewski1987} to determine the radial profiles of surface brightness, position angle and ellipticity of the cluster. Once this is done, the sky level is estimated for each photometric band independently, as the mean of the outer regions of the surface brightness profile. This constant value is then subtracted.

For consistency with the observational analysis, we applied the same masking procedure to the Horizon-AGN and Hydrangea mock images, to remove contaminating light from cluster member galaxies. %The simulation analysis was restricted to within 300 kpc of the BCG centre, and no residual sky background was included or subtracted.

\subsubsection{Isophotal fitting}
\label{subsec:isofit}

Following sky subtraction and  masking, \citet{kluge2020structure} derived $g$ band surface brightness ($\mu_g$), position angle, and ellipticity profiles using the \texttt{ellfitn} code \citep{bender1987morphological}. In this work, we use the publicly available versions of these profiles\footnote{Data available at \href{https://homepages.usm.uni-muenchen.de/mkluge/photometry.html}{https://homepages.usm.uni-muenchen.de/mkluge/photometry.html}}. 

To derive the surface brightness, PA, and $\epsilon$ profiles in the $g$, $r$, and $i$ bands for the BCG+ICL systems in the HSC dataset, we use the methodology described in Canepa et al. (in prep.). After applying the masking and sky subtraction procedures described in Section~\ref{subsec:skysub}, we fit elliptical isophotes using the \texttt{Photutils} implementation of the algorithm of \citet{Jedrzejewski1987}.
Uncertainties on the surface brightness profiles are computed following \citet{Golini2024} as the quadrature sum of the sky background error and the annular measurement error. 

Following Canepa et al. (in prep.), we define the maximum reliable radial extent of each HSC profile as the radius at which the measured surface brightness reaches the empirically determined background limited surface brightness threshold. This threshold, at $29-30$ mag/arcsec$^{2}$ depending on aperture size, is derived from random annulus measurements of PSF subtracted images. Consequently, our analysis is restricted to radii $<200$ kpc. This outer radius limitation is consistent with previous HSC ICL studies based on individual clusters without performing image stacking \citep[e.g.,][]{montes2021buildup, martinez2023galaxy}. The error on the ellipticity and position angle profiles are provided by the \texttt{Photutils} ellipse routine (see more details in \citealp{busko1996error}).

As for the K20 sample, we correct the $g$, $r$ and $i$ surface brightness profiles for cosmological dimming, K-correction and dust extinction.

By analysing profiles derived from two independent observational datasets and reduction pipelines, we aim to to assess whether the inferred inflection properties are intrinsic to the BCG+ICL systems rather than artefacts of the methodology.

We also apply the method of Canepa et al. (in prep.) to the mock simulated data from Hydrangea and Horizon-AGN cosmological simulations. 

The impact of spatial resolution is important when comparing systems across redshift, as lower-redshift clusters are sampled at smaller physical scales than their higher-redshift counterparts. To mitigate this effect, all profiles are resampled to have a common radial spacing of $1\,\mathrm{kpc}$ per bin, for both observational and simulated datasets\footnote{Considering a pixel scale of 0.168$\arcsec$/pix, 1 kpc corresponds approximately to 1 pixel at $z = 0.4$ and 3 pixels at $z = 0.1$.}.
This guarantees a consistent physical resolution across the sample. The effect of the binning choices on our results are discussed in Appendix~\ref{app:resolution}.

Finally, we assess the potential impact of PSF scattering whose main effect is to redistribute light from the central regions, leading to an artificial brightening of the outskirts \citep{2001trujillopsf, Sandin2014,2015SANDIN}. The effect depends on the instrument and wavelength used \citep{2014Abrahamvd,2015Duc,2015mihos}.
In the case of the BCG+ICL systems used in this study, light from the central regions of the BCG is redistributed to larger radii, mimicking the ICL contribution. Additional contamination can also arise from the extended PSF wings of nearby bright sources. While these two effects can influence the amount of diffuse light detected and thus affect the surface brightness levels, they are not expected to introduce or erase a structural transition in the profiles \citep[see e.g.,][]{Golini2025}. Therefore we do not proceed to remove the PSF contribution from the HSC dataset. Further details on the influence of the PSF effects on the determination of the transition radius are discussed in Appendix~\ref{app:effectofpsf}.

\section{Methods}
\label{sec:Methods}

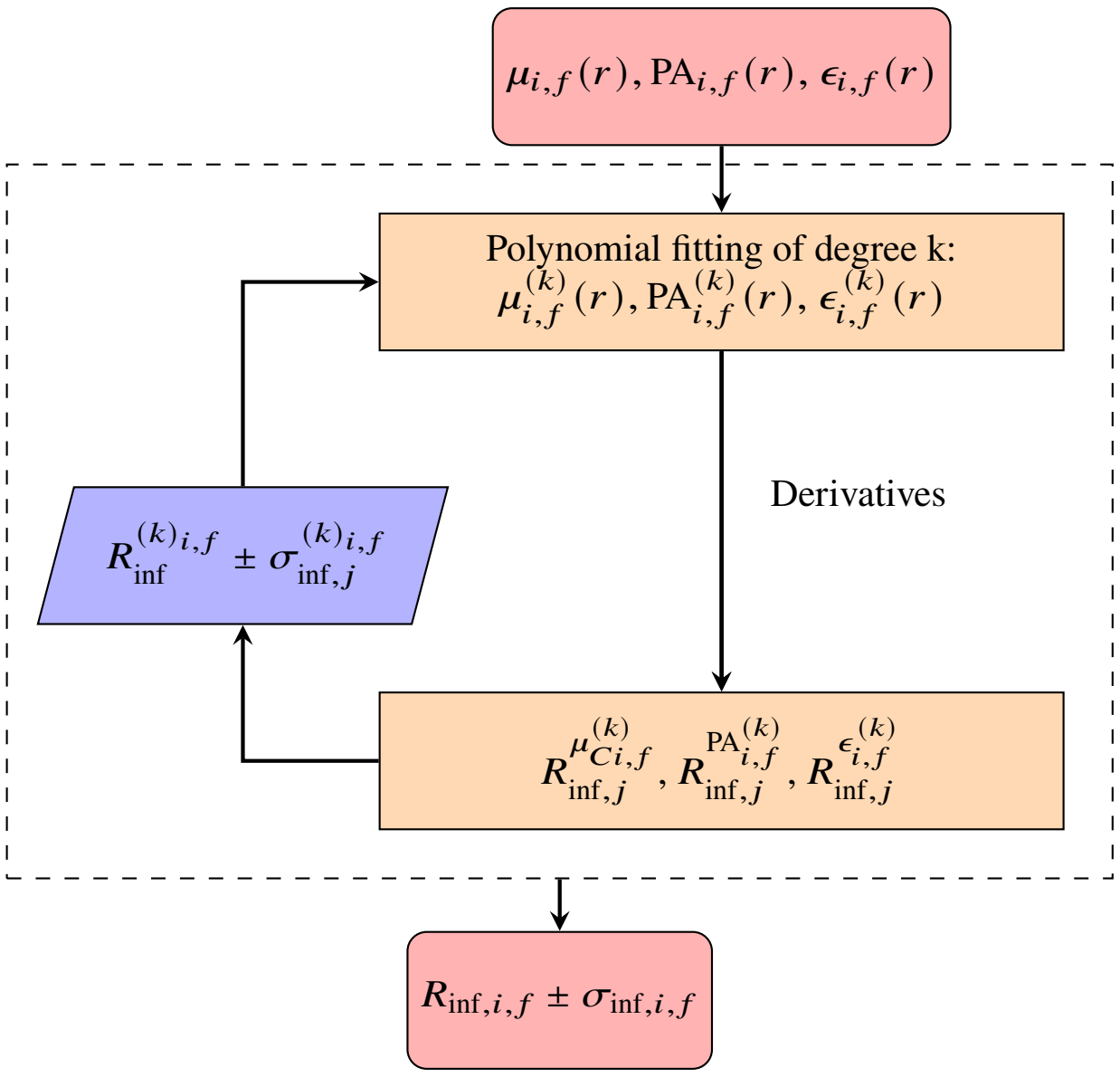
\begin{figure}
  \centering    
    \begin{tikzpicture}[node distance=2.5cm]
      \tikzstyle{startstop} = [rectangle, rounded corners, 
      minimum width=1cm, minimum height=1cm,
      text centered, draw=black, fill=red!30]
      
      \tikzstyle{io} = [trapezium, 
      trapezium stretches=true, trapezium left angle=70, 
      trapezium right angle=110, minimum width=3cm, minimum height=1cm, text centered, draw=black, fill=blue!30]
      
      \tikzstyle{process} = [rectangle, 
      minimum width=5cm, minimum height=1cm, align=center,
      draw=black, fill=orange!30]
      
      \tikzstyle{decision} = [diamond, 
      minimum width=3cm, minimum height=1cm, text centered, 
      draw=black, fill=green!30]
      \tikzstyle{arrow} = [thick,->,>=stealth]

      %\node (start) [startstop] { $ {i,f} $};
      \node (in1) [startstop] {$\mu_{i,f}(r), \mathrm{PA}_{i,f}(r), \epsilon_{i,f}(r)$};

      \node (pro1) [process, below of=in1, yshift=1cm] {Polynomial fitting of degree k: \\ $\mu^{(k)}_{i,f}(r), \mathrm{PA}^{(k)}_{i,f}(r), \epsilon^{(k)}_{i,f}(r)$};

      \node (pro1a) [process, below of=pro1,yshift=-1cm] {$R_{\mathrm{inf}, j}^{\mu^{(k)}_{Ci,f}}$, $R_{\mathrm{inf}, j}^{{\rm PA}^{(k)}_{i,f}}$, $R_{\mathrm{inf}, j}^{\epsilon^{(k)}_{i,f}}$};
      
      \node (out1) [io, left of=pro1a, xshift=-1cm, yshift=1.5cm] {{ {$R_{\mathrm{inf}}^{{(k)}_{i,f}} \pm \sigma_{\mathrm{inf}, j}^{{(k)}_{i,f}}$}}};
      
      %\draw [arrow] (start) -- node[anchor=south] {Isophote fitting} (in1);
      \draw [arrow] (in1) -- (pro1);
      \draw [arrow] (pro1) -- (pro1a);
      \draw [arrow] (pro1a) -| (out1);
      \draw [arrow] (out1) |- (pro1);

      \node (groupbox) [rectangle,draw,dashed, inner sep=10pt,fit=(pro1) (pro1a) (out1)] {};

      \node (out2)[startstop, below of = groupbox, yshift=-1cm] {$R_{\mathrm{inf}, {i,f}}\pm \sigma_{\mathrm{inf}, {i,f}}$};
      \draw [arrow] (groupbox) -- (out2);
      \draw [arrow] (pro1) -- node[anchor=south, xshift=1cm] {Derivatives} (pro1a);   
    
    \end{tikzpicture}
    
  \caption{Flowchart illustrating the main steps adopted in this work to detect the inflection radius ($R_{\mathrm{inf},i,f}$) for a cluster, $i$, in the photometric band $f$ and corresponding uncertainty.}
  \label{flowchart}
\end{figure}

To identify structural transitions in the radial profiles, we adopt the concept of inflection points, defined as locations where the second derivative of a function changes sign (equals to 0), corresponding to a change in curvature. In BCG+ICL systems, the radial profiles of $\mu$, PA, and $\epsilon$ are expected to vary smoothly with radius, but may exhibit changes in curvature associated with underlying structural variations.
Inflection points therefore provide a consistent way to characterise such variations using different observables. We exploit this concept to define whether there are common transition radii, defined by common inflection radii between $\mu$, PA, and $\epsilon$ profiles.
When detected, we define this characteristic scale as the inflection (or transition) radius, $R_{\mathrm{inf}}$, adopting this notation to distinguish it from similar definitions used in the literature (e.g., r$_{\times}$ in \citealp{kluge2021photometric}, R$_{tr}$ in \citealp{ragusa2021vegas,ragusa2022intra,ragusa2023does}, R$_{transition}$ in \citealt{Contini2022} or $R_{trans}$ in \citealt{Contini2024}).
To guide the reader through the text, a schematic overview of the analysis workflow is presented in Fig.~\ref{flowchart} and described below.

In the following, for each cluster, $i$, in the photometric band $f$, we will define $R_{\mathrm{inf},i,f}$ derived from the set: 
\begin{equation}
  \label{eq:profiles}
  \mu_{i,f}(r), \mathrm{PA}_{i,f}(r), \epsilon_{i,f}(r)
\end{equation}
Given the discrete radial sampling of the profiles, changes in curvature cannot be directly identified without interpolation. To facilitate this, we approximate each profile with a polynomial (via \texttt{Polynomial.fit}). For consistency, the three profiles in Eq.~\ref{eq:profiles} are interpolated using polynomials of the same degree $k$:
\begin{equation}
  \label{eq:poly}
  \mu^{(k)}_{i,f}(r), \mathrm{PA}^{(k)}_{i,f}(r), \epsilon^{(k)}_{i,f}(r).
\end{equation}

From these smooth representations, we evaluate the first and second derivatives (using \texttt{np.gradient}) in order to identify locations where the curvature changes sign. These locations correspond to a number j of inflection points for each radial profile, which we denote:
\begin{equation}
\label{eq:Rinf}
R_{\mathrm{inf}, j}^{\mu^{(k)}_{i,f}}, \;
R_{\mathrm{inf}, j}^{\rm PA^{(k)}_{i,f}}, \;
R_{\mathrm{inf}, j}^{\epsilon^{(k)}_{i,f}}
\end{equation}

We then examine whether the three profiles exhibit a coincident inflection within a tolerance  $\sigma_{\mathrm{inf}, j}^{{(k)}_{i,f}}$ $<$ 5 kpc. 
When such a coincidence is identified, if the values of the individual inflections $R_{\mathrm{inf}, j}^{\mu^{(k)}_{i,f}}$,
$R_{\mathrm{inf}, j}^{{\rm PA}^{(k)}_{i,f}}$$
R_{\mathrm{inf}, j}^{\epsilon^{(k)}_{i,f}}$ are not exactly the same, we keep their mean value as
$R_{\mathrm{inf}}^{{(k)}_{i,f}}$. In cases where multiple candidates are present, we retain the one with the smallest associated uncertainty.
%%%%%%% chosing polynomials
To ensure that this identification is also not driven by a particular choice of polynomials, the analysis is repeated over a range of polynomial degrees, from $k = 28$ (sufficient to capture the main structures in the profiles) up to $k = 34$, above which overfitting becomes significant. This results in a set of estimates which are combined into a single representative value using an inverse variance weighted mean:
\begin{equation}
R_{\mathrm{inf}, {i,f}} = \frac{\sum^{34}_{k=28} R_{\mathrm{inf}}^{{(k)}_{i,f}} / (\sigma_{\mathrm{inf}}^{{(k)}_{i,f}})^2}{\sum^{34}_{k=28} 1 / (\sigma_{\mathrm{inf}}^{{(k)}_{i,f}})^2}
\end{equation}

with corresponding uncertainty:
\begin{equation}
\sigma_{\mathrm{inf}, {i,f}} = \sqrt{\frac{1}{\sum^{34}_{k=28} (\sigma_{\mathrm{inf}}^{{(k)}_{i,f}})^2}} \times \sqrt{\chi^2_{\rm red}}
\end{equation}
where the Birge factor $\sqrt{\chi^2_{\rm red}}$ \citep{2014Birge} accounts for possible error underestimation.

\begin{figure*}
  \includegraphics[width=\linewidth]{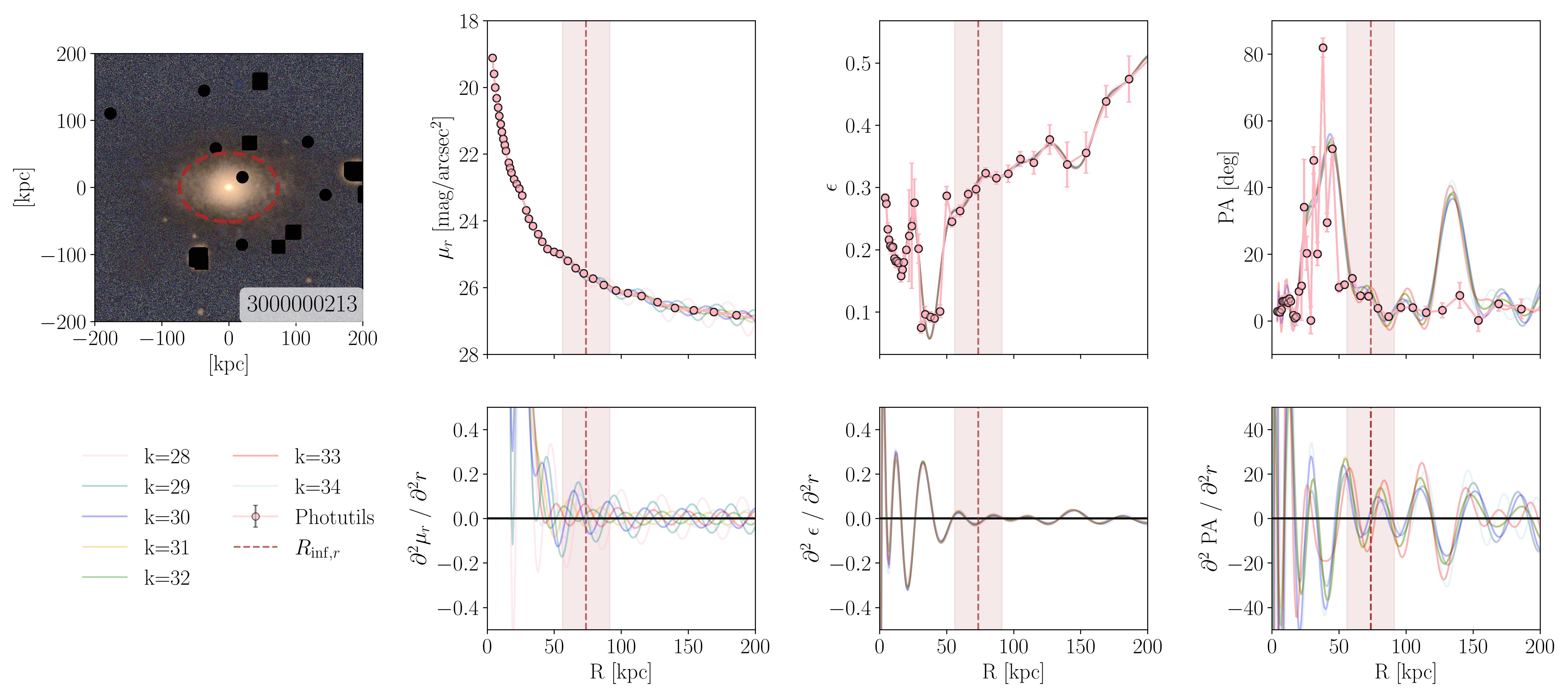}
  \caption{Determination of the inflection radius for cluster 3000000213 in the Hydrangea simulation. The top row shows the masked mock image and the polynomial fits to the $r$  band surface brightness, ellipticity and position angle profiles, which were derived using \texttt{Photutils} (dots). The bottom row presents the second derivative profiles of the polynomial fits. The vertical dashed lines in the top panels, and the shaded regions denote $R_{\mathrm{inf},r} \pm \sigma_{\mathrm{inf},r}$. The dashed ellipse in the masked image indicates the transition radius. %The diamonds mark the inflection radii for each polynomial fit.
  }
  \label{fig:ex}
\end{figure*}

The resulting $R_{\mathrm{inf}, {i,f}} \pm \sigma_{\mathrm{inf}, {i,f}}$ is taken as the inflection radius for the cluster $i$ in the photometric band $f$. We also denote the surface brightness value at this radius as $\mu(R_{\mathrm{inf}, {i,f}})$.
To help the reader understand the methodology described in Sect.~\ref{sec:Methods}, we provide a visual representation of it in Fig.~\ref{fig:ex}. The top-left panel shows a colour image of the mock image of cluster 3000000213 from the Hydrangea simulations, redshifted to $z = 0.1$ derived in Sect.~\ref{subsec:Mockimages}. The grayscale image is constructed by combining the $g$ , $r$  , and $i$ band mock images using the task \textit{astscript-color-faint-gray} \citep{scriptcolorimagesRaul}. Masked sources are indicated.

The remaining panels in the top row of Fig.~\ref{fig:ex} present polynomial fits (of degrees 28 to 34) to the extracted radial profiles in the $r$  band: surface brightness (left), ellipticity (centre), and position angle (right). The bottom row shows the corresponding second derivatives of these polynomial fits. The adopted transition radius for this cluster in the $r$ band is indicated by a vertical dashed line. Shaded regions represent the associated uncertainties. The ellipse corresponding to the inflection point is shown in the top left panel.
We remark here, that the impact of spatial sampling and smoothing on our measurements is presented in Appendix \ref{app:resolution}.

In summary, for each cluster and for each polynomial of degree $k$, we measured inflection radii independently of the surface brightness, ellipticity and position angle profiles. We selected the radii that agreed within a maximum of 5 kpc and adopted their mean as the inflection radius for that degree of polynomial. The final inflection radius for each cluster was obtained by taking the weighted average of the measurements derived from polynomial fits of degrees 28 to 34.

\section{Results}
\label{results}

The above procedure is applied independently to each cluster in all datasets (K20, HSC, Horizon-AGN and Hydrangea) and repeated separately for each available photometric band ($f = g$ for K20, and $f = g, r, i$ for the remaining datasets). In the following, we present the results organised by dataset and investigate potential correlations between the inferred transition radii and cluster properties. The mean values derived for each dataset are summarised in Table~\ref{Table:results} in Appendix~\ref{app:table}.

\subsection{Location of the Inflections}
\label{resu:location}

\begin{figure}
  \centering
  \includegraphics[width=\linewidth]{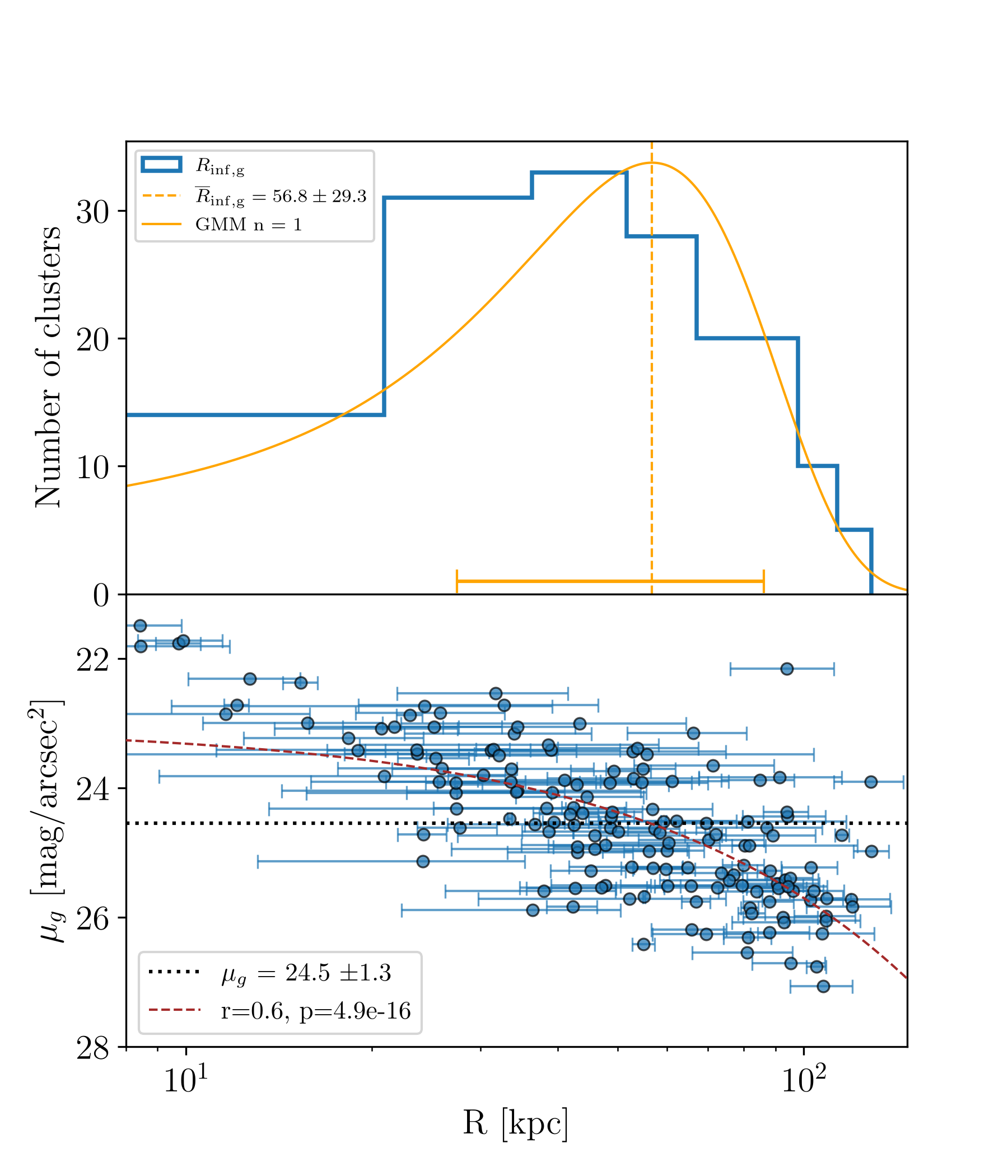}
  \caption{Top panel: Distribution of the characteristic radius $R_{\mathrm{inf}({i,g})}$ for the cluster sample in K20. The vertical dashed line and error bar indicate the 3$\sigma$-clipped mean and the corresponding standard deviation is shown as a horizontal bar. Overlaid is a one component GMM fit, scaled to the histogram counts. Bottom panel: Surface brightness $\mu(R_{\mathrm{inf}{,i,g}})$ as a function of $R_{\mathrm{inf}{,i,g}}$, with corresponding uncertainties. The dashed line represents a linear fit to the data. The mean of the surface brightness distribution is displayed as a horizontal dotted line.}
  \label{fig:hist-kluge}
\end{figure}

\textit{K20}: 
For all 170 systems in the K20 sample, we identify evidence for a common inflection radius in the $g$ band profiles of $\mu$, PA, and $\epsilon$. 
The top panel of Fig.~\ref{fig:hist-kluge} shows the distribution of the corresponding $R_{\mathrm{inf},g}$ values, using Freedman–Diaconis binning. The dashed vertical line and horizontal bar indicate the 3$\sigma$-clipped mean and its associated uncertainty, $\overline{R}_{\mathrm{inf},g} = \meanKluge \pm \sigmameanKluge$ kpc.
To guide the interpretation of the distribution, we overplot a single-component Gaussian Mixture Model (GMM), scaled to match the histogram counts.
The bottom panel of Fig.~\ref{fig:hist-kluge} shows the relation between $R_{\mathrm{inf},i,g}$ and the corresponding surface brightness $\mu(R_{\mathrm{inf},i, g})$, including uncertainties. The dashed line indicates the best fitting linear relation showing a moderate Pearson \citep{benesty2009pearson} linear correlation (r = 0.6, p = 4.9e-16). We therefore observe that, larger transition radii are associated with fainter surface brightness levels, ranging from $\sim 21$ mag/arcsec$^{2}$ to $\sim$ 28 mag/arcsec$^{2}$. This behaviour is consistent with expectations, as the surface brightness (and corresponding mass distribution) decreases with increasing radius from the central BCG. The mean $g$ band surface brightness at the transition radius between all the clusters is $\mumeanKluge \pm \sigmamumeanKluge$ mag/arcsec$^{2}$ (dotted horizontal line in the bottom panel of Fig.~\ref{fig:hist-kluge}).
Similarly to Fig.~\ref{fig:hist-kluge}, we report in Fig.~\ref{fig:kluge1sersic} and Fig.~\ref{fig:kluge2sersic} in Appendix \ref{app:klugehist12}, the histograms of the distribution of the inflection radii and corresponding surface brightness for the single Sérsic (117 clusters) and double Sérsic samples (47 clusters) defined in \cite{kluge2020structure} respectively.
We note that, while inflections are detected in all systems, their prominence may not be uniform across the sample. In particular, we might expect stronger inflections in systems with a more distinct two component (BCG+ICL) structure, whereas systems well described by a single Sérsic profile may exhibit more subtle features. Assessing this quantitatively is beyond the scope of the present work and is left for future work.

\begin{figure*}
  \includegraphics[width=\linewidth]{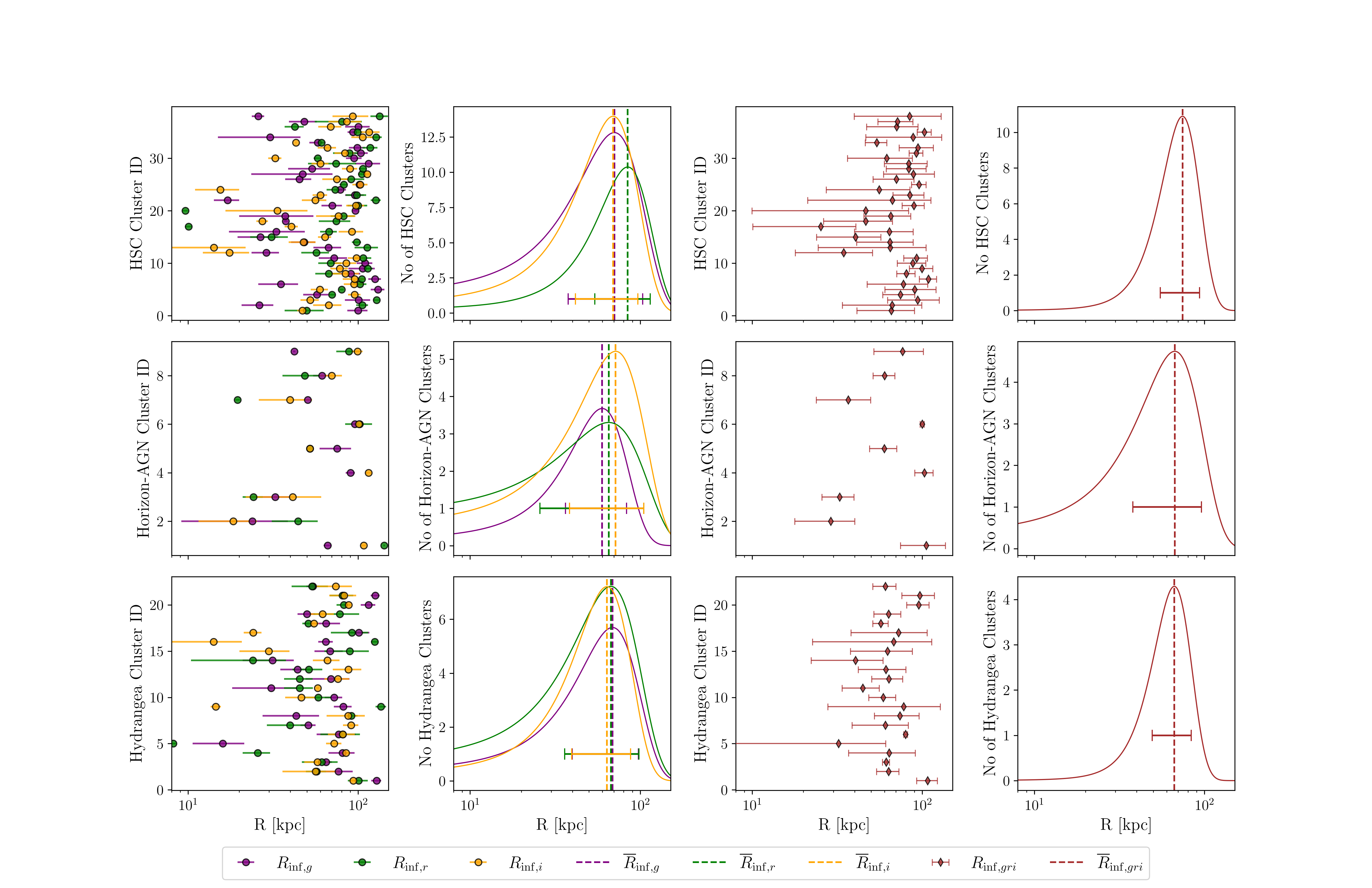}
  \caption{Comparison of the characteristic inflection radii for the HSC, Hydrangea and Horizon-AGN cluster samples. Left: Individual transition radius measurements as a function of cluster ID. Circles denote the mean radius obtained for different bands and corresponding uncertainties. Centre left: GMM distributions of the transition radii for $R_{\mathrm{inf},g}$ , $R_{\mathrm{inf},r}$ , and $R_{\mathrm{inf},i}$. Dashed vertical lines indicate the mean values. Centre right: Diamonds and horizontal error bars represent the mean radius and the corresponding standard deviation of each the three bands for each cluster ID. Right: Distribution of $R_{\mathrm{inf},gri}$, with the dashed line marking the mean radius. The corresponding uncertainty is shown as a horizontal line.}
  \label{fig:meanSIMULATIONS}
\end{figure*}

\textit{HSC}: The method described in Sect.~\ref{sec:Methods} is also applied to the HSC dataset in the $g$, $r$, and $i$ bands.
The upper left panel of Fig.~\ref{fig:meanSIMULATIONS} shows the transition radius measured in the three photometric bands for each cluster. The central right panel shows the mean transition radius across the bands for each cluster, together with the corresponding standard deviation, indicated by the diamond symbols and horizontal error bars.
The central left panel on the same row illustrate kernel density estimates of the distributions of $R_{\mathrm{inf},g}$, $R_{\mathrm{inf},r}$, and $R_{\mathrm{inf},i}$. The means and associated uncertainties are marked by dashed vertical lines: $\overline{R}_{\mathrm{inf},g} = \meanhscg \pm \sigmameanhscg$ kpc, $\overline{R}_{\mathrm{inf},r} = \meanhscr \pm \sigmameanhscr$ kpc, and $\overline{R}_{\mathrm{inf},i} = \meanhsci \pm \sigmameanhsci$ kpc. The distributions are broadly consistent, although the $r$ band shows a slight shift toward larger radii.

To assess whether the inflection radii are consistent across different photometric bands for individual clusters, we compared the transition radii measured in the $g, r$, and $i$ bands. We find that approximately 60$\%$ of the objects have $\overline{R}_{\mathrm{inf},g}$ and $\overline{R}_{\mathrm{inf},r}$ values consistent within 30 kpc, while about 50$\%$ show the same level of agreement between the $g$ and $i$ bands. Overall, we do not expect strong differences in the inferred transition radii across optical filters, since these bands trace similar stellar components. However, offsets may arise because the $g$ band is more sensitive to recent star formation, whereas the $i$ band is dominated by older stellar populations. Discrepancies may also result from the selection of multiple candidate inflection points in individual cluster profiles. A detailed investigation of these effects is beyond the scope of the present work and left for future study using a larger set of photometric filters and data (such as LSST), as discussed in Section \ref{sec:discussion}.
Therefore, in the following, we simply compute, for each cluster, the average transition radius across the three bands, resulting in a unimodal distribution with mean $\overline{R}_{\mathrm{inf},gri} = \meanhscgri \pm \sigmameanhscgri$ kpc (see the upper right panel of Fig.~\ref{fig:meanSIMULATIONS}).
The average surface brightness values at the transition radius are $\mumeanhscg \pm \sigmamumeanhscg$ mag/arcsec$^{2}$, $\mumeanhscr \pm \sigmamumeanhscr$ mag/arcsec$^{2}$, and $\mumeanhsci \pm \sigmamumeanhsci$ mag/arcsec$^{2}$ in the $g$, $r$, and $i$ bands, respectively. These values are consistent with previous HSC-based studies of the ICL \citep[e.g.,][]{montes2021buildup, martinez2023galaxy}. The full distribution of surface brightness values at the transition radius is shown in Fig.~\ref{fig:hist_ZeroCrossings_all} of Appendix~\ref{app:hist-individual bands}.

\textit{Hydrangea and Horizon-AGN}: 
For the cosmological simulations, we are able to identify a common structural transition radius for all 24 Hydrangea clusters and 7 Horizon-AGN clusters, using the radial profiles of $\mu$, PA, and $\epsilon$ derived from the mock images (see Sect.~\ref{subsec:Mockimages}) in the $g$, $r$, and $i$ bands. The results are shown in the central (Horizon-AGN) and bottom (Hydrangea) rows of Fig.~\ref{fig:meanSIMULATIONS}, following the same layout adopted for the HSC data.
For Horizon-AGN, the mean transition radii and corresponding uncertainties are $\overline{R}_{\mathrm{inf},g} = \meanhorizong \pm \sigmameanhorizong$ kpc, $\overline{R}_{\mathrm{inf},r} = \meanhorizonr \pm \sigmameanhorizonr$ kpc, and $\overline{R}_{\mathrm{inf},i} = \meanhorizoni \pm \sigmameanhorizoni$ kpc, yielding to an average value between bands of $\overline{R}_{\mathrm{inf},gri} = \meanhorizongri \pm \sigmameanhorizongri$ kpc. The distributions in the three bands are consistent within uncertainties, with the $r$  and $i$ band distributions in closer agreement with those measured with the HSC dataset. The $g$ band mean is rather smaller. 
For Hydrangea, the mean values of the distributions are $\overline{R}_{\mathrm{inf},g} = \meanhydrangeag \pm \sigmameanhydrangeag$ kpc, $\overline{R}_{\mathrm{inf},r} = \meanhydrangear \pm \sigmameanhydrangear$ kpc, and $\overline{R}_{\mathrm{inf},i} = \meanhydrangeai \pm \sigmameanhydrangeai$ kpc. These values are very similar across bands and exhibit a smaller scatter than observed in both HSC and Horizon-AGN. The band-averaged transition radius is $\overline{R}_{\mathrm{inf},gri} = \meanhydrangeagri \pm \sigmameanhydrangeagri$ kpc.
As for the HSC dataset, the surface brightness values at the transition radius for each cluster, for both simulation datasets are shown in Fig.~\ref{fig:hist_ZeroCrossings_all} (Appendix~\ref{app:hist-individual bands}). On average, these values are brighter than their observational counterparts, ranging between $ \mumeanhorizoni \pm \sigmamumeanhorizoni$ mag/arcsec$^{2}$ in the $i$ band (Horizon-AGN) to $ \mumeanhydrangeag \pm \sigmamumeanhydrangeag$ mag/arcsec$^{2}$ in the $g$ band (Hydrangea).

%%%%%%%%%%%%%%%%%%%%%%%%%%%%%%%%%%%%%%%%%%%%%%%%%%%%%%%%%%%%%%%%%%%%%%%%

\subsection{Inflections and host cluster properties}
\label{subsec:correlations}

In the following, we investigate the relationship between the inflection radii and the global properties of the host clusters. 

\begin{figure}
  \centering
  \includegraphics[width=\linewidth]{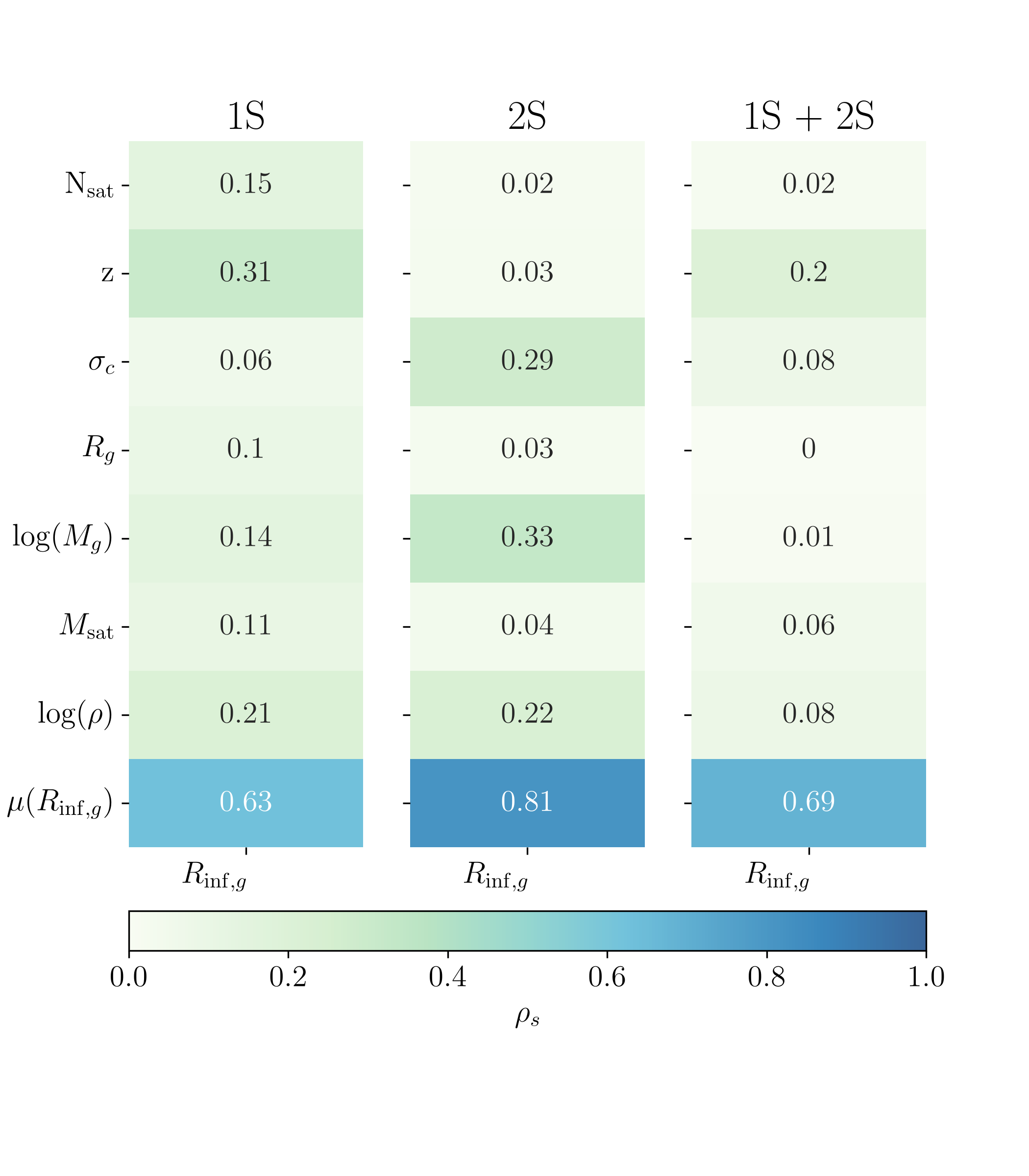}
  \caption{Correlations between transition radius for the single Sérsic (1S), double Sérsic (2S) and 1S +2S samples (horizontal axis) and cluster parameters (vertical axis). The rows show number of satellite galaxies, redshift of the cluster, the velocity dispersion of the satellite galaxies, gravitational radius, gravitational mass, integrated absolute brightness of all satellite galaxies (excluding the BCG+ICL), log($\rho$) and $\mu (R_{\mathrm {inf,g}})$. The spearman coefficient for each correlation is given in the tiles. }
  \label{fig:correlationsKluge}
\end{figure}

\textit{K20:} For the \citet{kluge2020structure} dataset, the cluster properties are taken from Table~2 of \citet{kluge2021photometric}, noting that these measurements are only available for a subset of 38 systems. 
Fig.~\ref{fig:correlationsKluge} shows the spearman \citep{wissler1905spearman,myers2014spearman} correlation coefficients between the transition radius, and cluster properties for three different modeling approaches defined in \cite{kluge2020structure} (single Sérsic and double Sérsic). We use the spearman rank correlation coefficient to assess monotonic relationships between parameters, with $\rho = 1$ indicating a perfect correlation and $\rho_s = 0$ indicating no correlation.
The parameters considered include the velocity dispersion of satellite galaxies ($\sigma_c$ [km/s]; \citealp{lauer2014brightest}), richness (N$_{\rm sat}$; number of satellite galaxies), gravitational radius ($R_g$ [kpc]), integrated absolute brightness of all satellite galaxies ($M_{\rm sat}$ [g$^\prime$ mag]), logarithm of the gravitational mass (log($M_g$) [log(M$_{\odot}$)]) and logarithm of the mass density log($\rho$) [log(M$_{\odot}$/kpc$^3$)] \citep{kluge2021photometric}. Correlation coefficients are shown in the corresponding tiles.
In all configurations (single Sérsic, double Sérsic and the combination of the two), we find a strong but physically unsurprising correlation between $R_{\mathrm{inf},g}$ and the surface brightness at that location, $\mu(R_{\mathrm{inf},g})$ mag/arcsec$^2$, consistent with fainter surface brightness at larger radii ($\rho_s$ > 0.63). Beyond this relation, correlations with global cluster properties are weak. For the double Sérsic subsample, there is a $\rho_s$ = 0.33 correlation with $\log(M_g)$, which is not present in the single Sérsic case and negligible when the full sample is considered. 
There is also a $\rho_s$ = 0.29 correlation with $\sigma_c$ while the single Sérsic subsample correlates more with redshift ($\rho_s$ = 0.31).
Overall, these results indicate that the transition radius is more closely related to the distribution of the mass and light profile rather than to global cluster properties.
We note however, that the cluster properties (mass, luminosity, and radius) estimates of \citet{kluge2021photometric} are uncertain due to the simplified measurement approach and may mask weak correlations, which could be revealed through the use of more accurate cluster property determinations based on cluster-finding methods (for luminosities) and X-ray or lensing measurements (for cluster masses).
The lack of significant correlations in the K20 sample may also be attributable to the fact that the analysis is restricted to the $g$ band. We therefore extend the analysis to the HSC sample, which provides measurements in three photometric bands ($g$, $r$, and $i$).

\begin{figure}
  \centering
  \includegraphics[width=\linewidth]{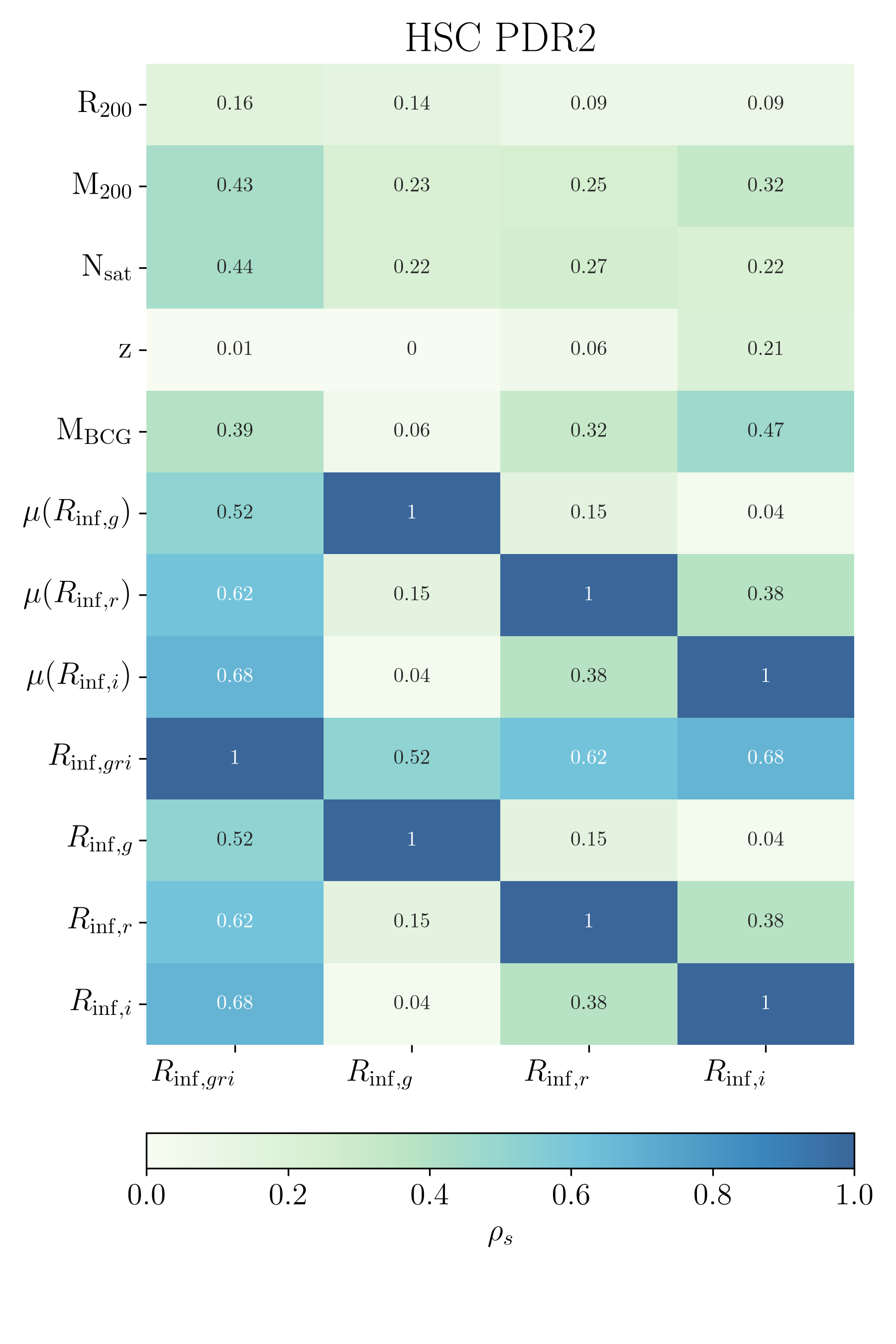}
  \caption{Full matrix showing the correlation strength between the mean inflection radius and various cluster properties for the HSC subsample. The specifics of the cluster properties are described in the text.}
\label{fig:corr_hsc}
\end{figure}

\textit{HSC:} As mentioned in Sect.~\ref{subsubsec:hsc}, in the HSC catalogue, BCGs are identified using the CAMIRA algorithm, which provides estimates of the BCG stellar mass, cluster richness, and photometric redshift. From the richness and redshift of each system, we further derive estimates of $\sigma_{\rm c}$, and $r_{200}$ following \citet{viola2015dark} and \citet{oliva2017study}. The resulting correlation coefficients between the inflection-derived quantities in the $g$, $r$, and $i$ bands and the cluster properties are shown in Fig.~\ref{fig:corr_hsc}.
Consistent with the results obtained for the K20 sample, the strongest correlations are found between the transition radius and the corresponding surface brightness at that radius in all three photometric bands ($\rho_s$ = 1) although this relation is driven by the underlying radial structure of the light profile. We observe a weak ($\rho_s$ $\sim$ 0.47) correlation between the transition radius, and M$_{\rm BCG}$ for the $i$ band. We find no strong dependence on dark matter related quantities, such as M$_{\rm 200}$ or R$_{200}$.

\begin{figure}
    \centering
    \includegraphics[width=\linewidth]{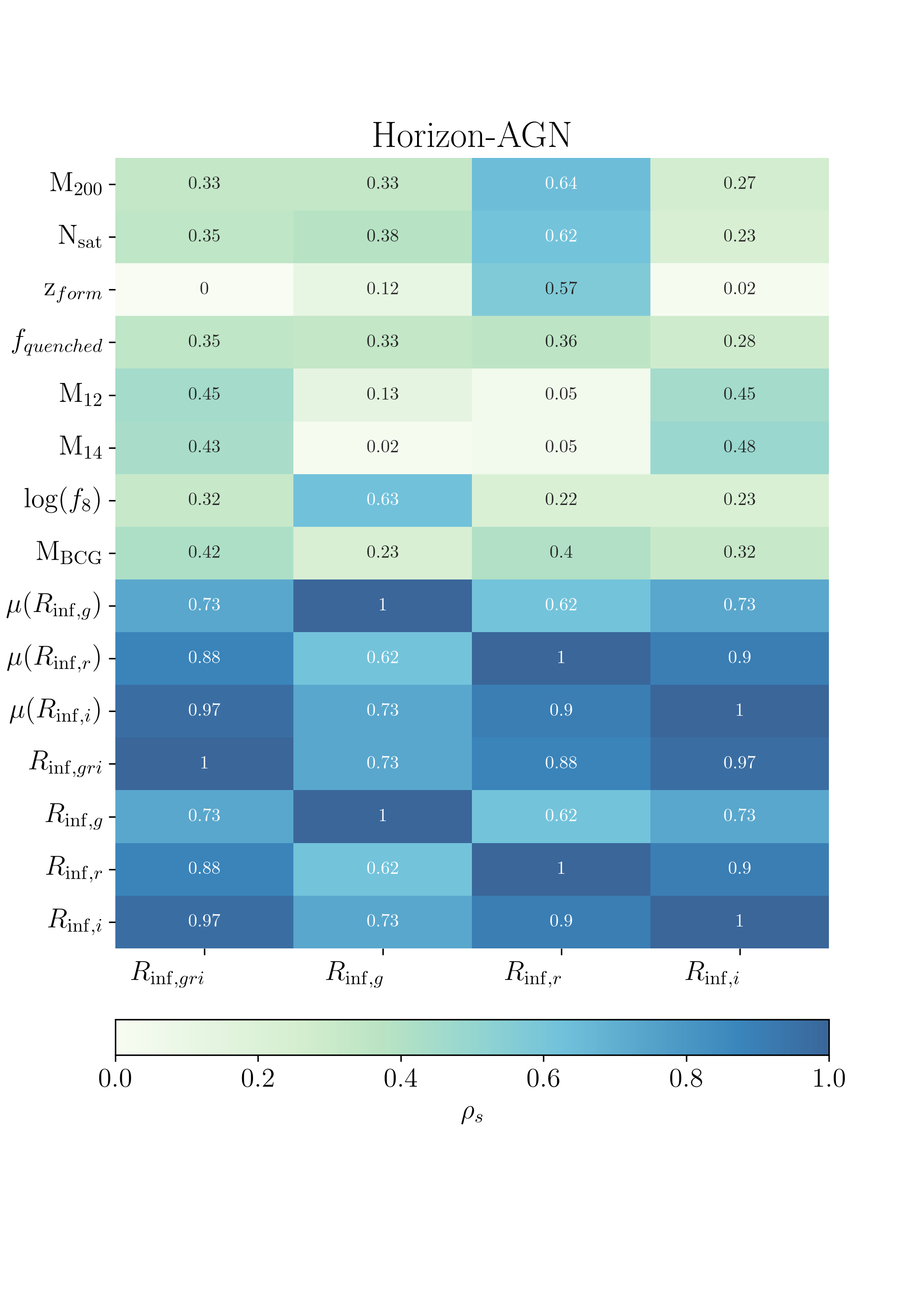}
        \caption{Full matrix showing the correlation strength between the mean inflection radius and various cluster properties for the Horizon-AGN simulated dataset. The specifics of the cluster properties are described in the text.}
    \label{fig:corr_simhorizon}
\end{figure}

\begin{figure}
        \centering
        \includegraphics[width=1\linewidth]{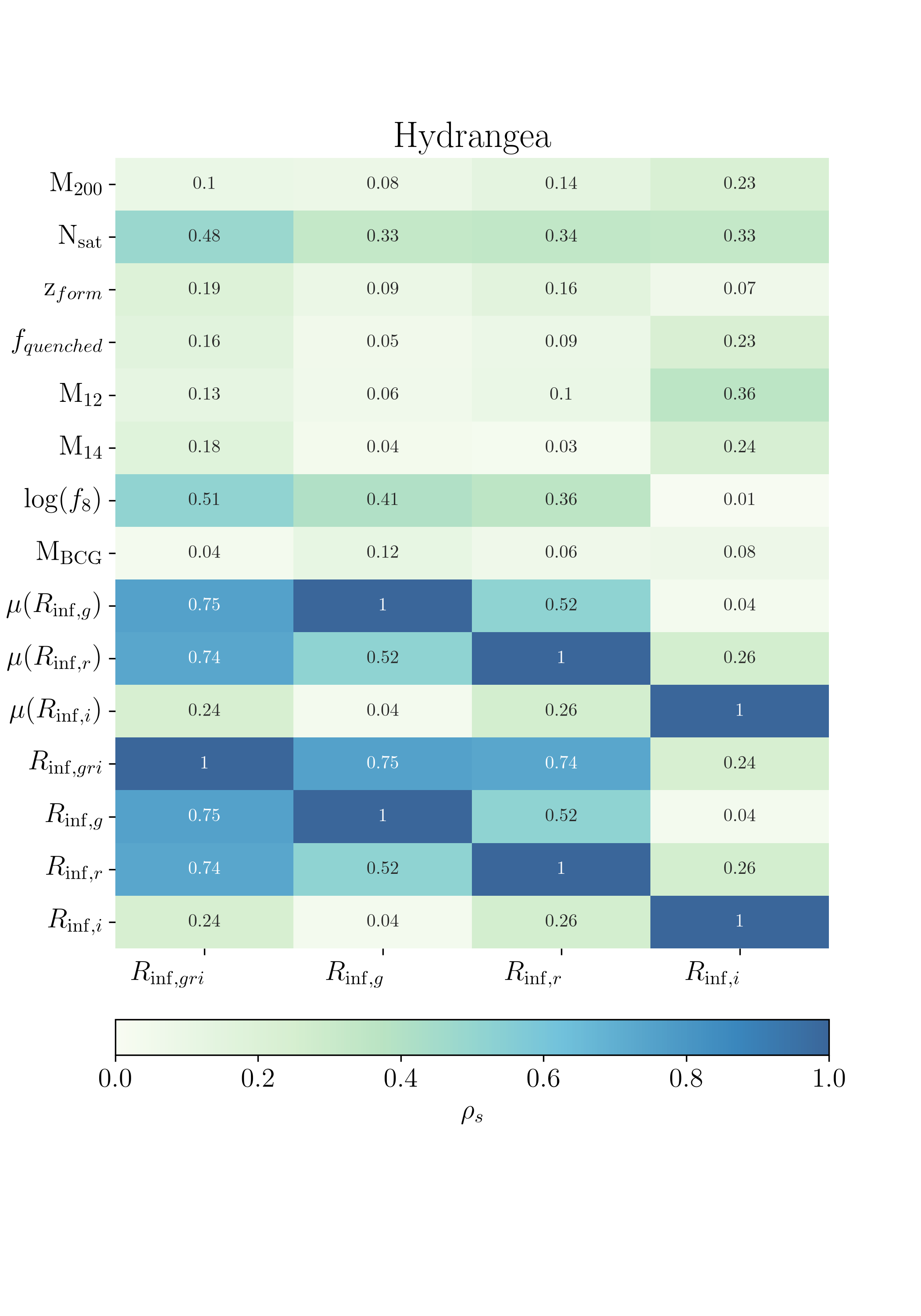}
        \caption{Similar to Fig.~\ref{fig:corr_simhorizon}, full matrix showing the correlation strength between the mean inflection radius and cluster properties for the Hydrangea dataset. The specifics of the cluster properties are described in the text.}
        \label{fig:corr_simhydrangea}
    \end{figure}

%%%%%%%%%%%%%%%%%%%%%%%%%%%%%%%%%%%%%%%%%%%%%%%%%%%%
\textit{Hydrangea and Horizon-AGN:} Fig.~\ref{fig:corr_simhorizon} and Fig.~\ref{fig:corr_simhydrangea} show the correlation coefficients between the transition radius properties and host cluster properties for the Horizon-AGN and Hydrangea simulations, respectively\footnote{Cluster properties for the simulated samples are publicly available at \href{https://www.nottingham.ac.uk/astronomy/NottICL/cluster_mocks/}{https://www.nottingham.ac.uk/astronomy/NottICL/cluster\_mocks/}}.
For both simulation suites, the same set of parameters displayed along the rows and columns of the correlation matrices. These include the halo mass M$_{200}$, the number of satellites with M$_\star > 10^{10} {\rm M_\odot}$ (N$_{\rm sat}$), the halo formation redshift (z$_{ form}$), the quenched fraction of satellite galaxies $f_{quenched}$, the stellar mass ratio between the BCG and the second- and fourth-most massive galaxies within 0.1R$_{200\rm c}$ (M$_{12}$ and M$_{14}$, respectively), the logarithmic fraction of the total halo mass contained in the eighth most massive substructure (log($f_8$)) and the stellar mass of the BCG (M$_{\rm BCG}$). The surface brightness value at the location of the inflection is also reported for each photometric band.

For the Horizon-AGN sample, as in the observational samples, the strongest correlation is found between the transition radius and the surface brightness measured at that radius. This trend is present in all photometric bands and reflects the underlying radial distribution of stellar light ($\rho_s$ = 1).
Correlation between the inflection radii in the different photometric bands correlates with one another more than for the HSC counterpart ($\rho_s$ > 0.6).
The transition radius also exhibits a correlation ($\rho_s$ = 0.63) with log($f_8$) in the $g$ band, suggesting a connection between the extent of the transition region and the degree of substructure within the halo. 
The inflection in the $r$ band instead show a possible correlation with M$_{\rm 200}$, N$_{\rm sat}$ and z$_{form}$ that is weaker in the other bands ($\rho_s$ > 0.57).
A weaker correlation is also observed with M$_{12}$ and M$_{14}$ for the $i$ band ($\rho_s \sim 0.5$).

Despite the expected correlation between the location of the inflection and the corresponding surface brightness in the three photometric bands, the Hydrangea sample displays a somewhat different behaviour regarding the correlations with cluster properties. In this case, a weaker correlation is found with $f_8$ in the $g$ band ($\rho_s = 0.41$) and there is a trend between the transition radius in the average $gri$ and N$_{\rm sat}$ ($\rho_s = 0.48$). The transition radii measured in the $g$, $r$ and $i$ bands do not exhibit a strong mutual correlation, despite displaying very similar distributions.

Overall, the simulation results reinforce the observational finding that the transition radius is most strongly linked to the local surface brightness at the transition location, driven by the monotonic decrease of surface brightness with radius. Beyond this relation, no clear and consistent dependence emerges with global dark matter halo properties across both simulation suites. However, we note that this analysis does not rule out the connection between $R_{transition}$ and halo concentration found by \cite{Contini2022}.

\section{Discussion}
\label{sec:discussion}

We derived the radial profiles of the surface brightness, ellipticity and position angle of galaxy clusters using isophotal fitting and introduced an automated approach to locate inflection points. Across both the observed (K20 and HSC) and simulated datasets (Horizon-AGN and Hydrangea), we find a consistent alignment of inflection points in the surface brightness, ellipticity, and position angle profiles. This suggests evidence for a structural transition in the cluster properties.

\subsection{Physical interpretation of a common inflection}
\label{phys_interpret}

The coincidence of the inflection points in the surface brightness, ellipticity, and position angle profiles across our different and independent samples suggests that these observables trace a common structural transition within the BCG+ICL system. A physical motivation for this interpretation comes from dynamical studies of nearby clusters. Integrated-light spectroscopy and planetary-nebula kinematic measurements have shown that the stellar velocity dispersion increases with radius and, in the outer regions of the BCG, approaches that of the cluster galaxy population \citep[e.g.,][]{Dressler1979,Arnaboldi1996,bender2015structure,kelson2002determination,Longobardi2018}. These observations indicate a transition from a region dominated by the BCG potential to one increasingly influenced by the cluster potential.
This picture is also consistent with theoretical expectations for the assembly of the ICL and the dark matter halo. Both components are thought to grow through the accretion of material along preferred directions defined by the surrounding large-scale structure and cosmic filaments, and are therefore expected to retain information about the orientation of the cluster-scale potential \citep[e.g.,][]{rudick2006formation,kuchner2020mapping,castignani2022virgo,vurm2023cosmic}. By contrast, the inner BCG may reflect a different assembly history and different relaxation timescale and do not need to be perfectly aligned with the cluster halo, especially in systems that have experienced recent mergers.
Consequently, if the BCG and cluster-scale potentials differ in shape or orientation, the changing relative contribution of each component with radius can produce the observed variations in surface brightness, ellipticity, and position angle. The resulting PA twists and ellipticity changes are not inconsistent with the known correlation between BCG and cluster orientations \citep{binggeli1982shape,dubinski1998origin,west2017ten}; rather, they may reveal structural differences within an overall aligned system. In this context, our results are consistent with evidence that the ICL is more strongly aligned with the host cluster than the central BCG itself \citep{kluge2021photometric}, suggesting that the accreted stellar component retains a memory of the infall directions that shaped the cluster. The inflection points may therefore mark the radial scale at which the light distribution begins to transition from the relaxed, BCG-dominated core toward the cluster-aligned ICL component (unrelaxed).
%%%%%%%%%%%%%%%%%%%%%%%%%%%%%%%%%%%%%%%%%%%%%%%%%%%%%%%%%%

We remark that this interpretation does not imply the existence of a unique physical boundary between the BCG and the ICL. Instead, the observed transition may mark the radial range over which the relative contributions of these overlapping stellar components change most rapidly, reflecting the gradual shift from stars predominantly associated with the central galaxy to those more strongly coupled to the cluster environment.

To further highlight the strength of this methodology, we note its successful application in a different astrophysical context by \citet{Golini2025}. In that work, a similar second-derivative approach was used to identify structural changes in the stellar mass density profile of a Milky Way-like galaxy and to determine the outermost extent of \textit{in-situ} star formation \citep[see e.g.,][]{2020Trujillo,2022chamba-tracinglimits}. Especially in spirals, galaxy growth is primarily driven by the conversion of gas into stars, a process that is regulated by a critical gas density threshold \citep{1972Quirk, 2004Schaye, 2008rokroskar}: Once the gas reservoir falls below this threshold, star formation  ceases, leaving a characteristic break in the stellar mass density profile. The inflection point, identified through the second derivative of the radial mass profile, marks the radius at which the profile curvature reaches a minimum and the slope changes most abruptly, providing a physically meaningful estimate of the stellar "edge".
The transition between the BCG and the ICL differs from this scenario in that it is not expected to produce a sharp discontinuity. Instead, the BCG and ICL represent overlapping stellar components that merge gradually, resulting in a smooth structural transition rather than a distinct break. Nevertheless, the inflection point remains meaningful as it identifies the radius where the profile morphology changes most significantly and where the relative contribution of the two components begins to shift.

Beyond our interpretation, the transition may also mark the boundary of the BCG's "sphere of influence", separating the hierarchically assembled central galaxy from the surrounding intracluster material \citep{chen2022sphere}; alternatively, it may trace either the BCG's extended stellar halo, a dynamically mixed component linking the galaxy and the ICL \citep{Contini2018,Contini2022}, or a splashback-like feature in the stellar distribution associated with the orbital apocentres of accreted stars \citep{zhang2025splashback}.

\subsection{Transition radii across datasets}

The inferred transition radii identified in this work span $60-85$ kpc, with an intrinsic scatter of $20-40$ kpc across all datasets analysed. The K20 $g$ band mean measurement (\(\meanKluge \pm \sigmameanKluge\) kpc) is consistent, within the uncertainties, with the values obtained for the Hydrangea and Horizon-AGN simulations, as well as with those measured in the HSC sample ($\meanhydrangeag \pm \sigmameanhydrangeag$ kpc, $\meanhorizong \pm \sigmameanhorizong$ kpc and $\meanhscg \pm \sigmameanhscg$ kpc respectively). When all HSC filters are considered together, the transition radius is on average located at larger radii ($\meanhscgri \pm \sigmameanhscgri$ kpc) than in the simulation datasets. However, this offset remains small compared to the intrinsic scatter of the distributions. Although the average transition radii measured for the two observational (K20 and HSC) samples are not identical, they remain consistent within the associated uncertainties.

The observed offset may arise from a combination of differences in sample selection (e.g. halo-mass distribution), data depth, reduction procedures, and profile extraction methodologies. We therefore emphasise the consistent detection of the inflection feature across independent observational and simulation based datasets as the primary result, and regard its recurring detection as more robust than the absolute value of R$_{\rm inf}$ itself.

Despite the differences, all measurements fall within the broad range reported in the literature (30–200 kpc), with most observational estimates lying between 50 and 100 kpc, depending on depth and methodology \citep{Zibetti2005, Iodice2016,zhang2019dark, montes2021buildup,gonzalez2021discovery, contini2021brightest, ragusa2021vegas,ragusa2022intra,ragusa2023does,zhang2025splashback}. Focusing on the $g$ band alone, we observe that the K20 mean radius is in close agreement with that of Horizon-AGN. Among the datasets, the HSC exhibits larger transition radii in the $g$ band. 
HSC and Horizon-AGN mock images show a further transition radii in $i$ band with respect to Hydrangea. 

An important consistency emerges when considering the surface brightness associated with the transition radius. In HSC data, the mean inflection occurs at $\mu_g =\mumeanhscg \pm \sigmamumeanhscg$ mag/arcsec$^{2}$, $\mu_r =\mumeanhscr \pm \sigmamumeanhscr$ mag/arcsec$^{2}$ and $\mu_i =\mumeanhsci \pm \sigmamumeanhsci$ mag/arcsec$^{2}$ in agreement with the commonly adopted observational threshold of $\sim 26$–26.5 mag/arcsec$^{2}$ used to separate BCG and ICL \citep{feldmeier2004deep,Zibetti2005, montes2014intracluster, montes2021buildup, burke2015coevolution, kluge2021photometric, martinez2023galaxy, Presotto2014, furnell2021growth, Canepa2025}. This demonstrates that our method recovers the same distinction identified by classical surface brightness cuts.
In contrast, simulations (Hydrangea and Horizon-AGN) place the transition at brighter levels ($\sim 23$–24.5 mag/arcsec$^{2}$).  
We also note that the transition surface brightness measured for the K20 sample in the $g$ band ($\mu_g =\mumeanKluge \pm \sigmamumeanKluge$ mag/arcsec$^2$) is closer to the values inferred from the simulations than to those measured in the HSC data. However, this should not be over-interpreted. The origin of this offset remains uncertain and may reflect both physical and observational effects. On the simulation side, Hydrangea was not calibrated to reproduce the detailed surface brightness distributions of BCG+ICL systems, and the simulated BCGs are known to be systematically more massive than observed counterparts \citep{Bahe2017}. In addition, numerical resolution and sub-grid prescriptions may affect the amount and distribution of diffuse stellar material \citep[e.g.][]{cui2014characterizing, martin2024stellar}. On the observational side, the transition is measured at low surface brightness levels, where residual uncertainties associated with sky subtraction, source masking, cosmological dimming, extinction corrections, and K-corrections become increasingly important.

The datasets analysed here probe different redshift regimes: K20 at $z < 0.14$, simulations at $z = 0.1$, and HSC at $0.2 < z < 0.4$. 
While the larger mean R$_{\rm inf}$ measured in the HSC sample than in K20 could be suggestive of redshift dependent effects, the uncertainties are large and Fig.~\ref{fig:correlationsKluge} and Fig.~\ref{fig:corr_hsc} do not suggest a strong relationship. So, we cannot isolate redshift as the primary driver of the observed differences.
However, the presence of a well-defined transition radius across all probed redshifts is an important result in itself. This indicates that the transition between BCGs and ICLs is already in place over this range of cosmic time ($0 < z < 0.4$) and therefore represents a robust feature of cluster light profiles rather than a transient phenomenon.

Although a weak trend of cluster radius with cluster richness has been reported in DECaLS observations \citep{zhang2025splashback}, we only partially observe this relationship: In the Horizon-AGN sample, N$_{\rm sat}$ correlates more strongly in the $r$ band than in $g$ or $i$ (0.62, 0.38 and 0.23 respectively) while in Hydrangea the correlation in the three bands is closer to $\rho_s$ = 0.3, but cannot be clearly detected in the K20 or HSC samples. This suggests that dependence on global cluster properties may be sensitive to the specific dataset or methodology adopted.
The correlations linking transition features to cluster richness, halo mass, and dynamical state \citep{chen2022sphere, Contini2024,zhang2025splashback} are not consistently detected across all datasets in this work, highlighting the need for larger and more homogeneous samples.

\subsection{Future directions}

LSST will provide a large, homogeneous sample that will allow us to further investigate the presence of such a transition region, as well as any correlation with halo mass and redshift.
Also, the presence of a transition region in the combined BCG+ICL system suggests the existence of distinct stellar assembly channels. The inner regions are expected to be dominated by stars formed through hierarchical merging processes that build up the central BCGs 
over time \citep{de2007hierarchical,bezanson2009relation, oser2010two, brough2011spatial,wellons2016diverse,qu2017chronicle, pillepich2018simulating}, while the outer ICL regions are thought to be progressively enriched by tidally stripped material originating from satellite galaxies \citep{Rudick2009}. 
To support the hypothesis of different formation channels for BCG and ICL, observational studies of galaxy clusters have revealed clear radial colour gradients \citep[e.g.,][]{montes2014intracluster,demaio2015origin,iodice2017intracluster,mihos2017burrell,demaio2018lost,monte2018}, indicative of declining metallicity with radius. Moreover, the stellar populations of the ICL in nearby systems are found to be generally old \citep{williams2007metallicity,coccato2010distinct}. 
The next logical step is therefore to characterise the stellar populations at the transition radius. Upcoming LSST data, which will provide deep and homogeneous imaging in the $u, g, r, i, z$ bands, will enable the measurement of colour gradients across the BCG–ICL transition. 

In parallel, cosmological simulations provide an ideal framework for exploring the evolution of this feature. By analysing multiple snapshots and following the BCG along its merger tree, it will be possible to track the evolution of both the transition radius and the corresponding brightness with redshift. This will allow us to probe the growth of the ICL and the role of accretion processes in shaping the outer regions of galaxy clusters.

\section{Conclusions}
\label{conclusions}

Distinguishing between the central BCG and the extended ICL remains a significant challenge in the study of galaxy clusters, as there is no accepted definition to differentiate between the two components. Driven by the recent focus on the BCG–ICL transition region \citep[see e.g.,][]{ragusa2021vegas,chen2022sphere,ragusa2022intra,ragusa2023does, gonzalez2021discovery, Contini2022, Proctor2024,zhang2025splashback}, and supported by observational evidence suggesting that variations in surface brightness may coincide with changes in the geometric properties (PA, $\epsilon$) of the stellar distribution \citep{montes2021buildup,marrero2026large}, we explore the possibility of identifying a common structural behaviour in the radial profiles of surface brightness, ellipticity, and position angle of galaxy clusters.
To this end, we analysed observational data from \citet{kluge2020structure} and the HSC survey, as well as mock observations from the Horizon-AGN and Hydrangea cosmological hydrodynamical simulations. Using isophotal fitting, we extracted radial profiles of surface brightness, ellipticity, and position angle and we developed an automated method based on the identification of inflection points, which are defined as the zeros of the second derivative of each profile (see Figs.~\ref{flowchart} and \ref{fig:ex}). From this analysis, we conclude the following.

(i) Across all clusters in our datasets (see Fig.~\ref{fig:hist-kluge} and Fig.~\ref{fig:meanSIMULATIONS}), we find evidence for a common structural transition marked by coincident ($\sigma_{tol} < 5$ kpc) inflections in the radial profiles of surface brightness, ellipticity, and position angle. The presence of this feature suggest that the gradual transition between the BCG dominated component and the ICL is reflected in both the photometric distribution of stars and their geometric configuration. This builds on previous work that identified the transition from surface brightness profiles \citep[e.g.,][]{zhang2025splashback} by demonstrating that it is associated with consistent changes across multiple independent observables.
In this work, we define this feature at the radial location $R_{\rm inf}$ and, it is consistent between observational and simulation based datasets, occurring at $60-85$ kpc, with an intrinsic scatter of $20-40$ kpc. This result is in good agreement with earlier observational estimates \citep{montes2021buildup,gonzalez2021discovery,Iodice2016,yoo2021intracluster,spavone2024galaxy,bellhouse2025euclid} and theoretical predictions \citep{Contini2022}.

(ii) We find that larger transition radii correspond to fainter surface brightness levels %(bottom panels of Figs.~\ref{fig:hist-kluge} and \ref{fig:hist_ZeroCrossings_all})
, reflecting the monotonic decline of stellar density with radius in BCG+ICL systems. By contrast, we find no strong or consistent dependence of $R_{\rm inf}$ on global cluster properties across any of the observational or simulated datasets (Figs.~\ref{fig:correlationsKluge}, \ref{fig:corr_hsc}, \ref{fig:corr_simhorizon}, and \ref{fig:corr_simhydrangea}).

(iii) Our results indicate that the BCG-ICL transition is characterized by a coherent structural signature that appears simultaneously in surface brightness, ellipticity, and position angle profiles. The consistency of this feature across independent observations and cosmological simulations suggests that it traces an underlying change in the stellar distribution and therefore provides a robust and reproducible way to identify the transition. We emphasize that the second-derivative feature identified here should not be interpreted as a unique physical boundary between the BCG and ICL. Rather, it provides an observationally motivated definition of a characteristic transition scale that may be linked to the assembly history of the combined BCG+ICL system. A key advantage of this approach is that it avoids parametric decompositions and stacking, relying instead on a fully automated procedure applicable to individual clusters. This makes it particularly well suited to upcoming surveys such as Euclid \citep{laureijs2011euclid,scaramella2022euclid,bellhouse2025euclid} and LSST \citep{ivezic2019lsst}, which will enable systematic studies of how this transition relates to cluster assembly, stellar halo growth, and ICL formation across cosmic time.

\section*{Acknowledgements}
We thank the anonymous referee for their comments, which helped to improve the quality of the manuscript.
The Hyper Suprime-Cam (HSC) collaboration includes the astronomical communities of Japan and Taiwan, and Princeton University. The HSC instrumentation and software were developed by the National Astronomical Observatory of Japan (NAOJ), the Kavli Institute for the Physics and Mathematics of the Universe (Kavli IPMU), the University of Tokyo, the High Energy Accelerator Research Organization (KEK), the Academia Sinica Institute for Astronomy and Astrophysics in Taiwan (ASIAA), and Princeton University. Funding was contributed by the FIRST program from Japanese Cabinet Office, the Ministry of Education, Culture, Sports, Science and Technology (MEXT), the Japan Society for the Promotion of Science (JSPS), Japan Science and Technology Agency (JST), the Toray Science Foundation, NAOJ, Kavli IPMU, KEK, ASIAA, and Princeton University. 
This paper makes use of software developed for the Large Synoptic Survey Telescope. We thank the LSST Project for making their code available as free software at  http://dm.lsst.org.
This paper is based on data collected at the Subaru Telescope and retrieved from the HSC data archive system, which is operated by the Subaru Telescope and Astronomy Data Center (ADC) at NAOJ. Data analysis was in part carried out with the cooperation of Center for Computational Astrophysics (CfCA), NAOJ.
The Pan-STARRS1 Surveys (PS1) have been made possible through contributions of the Institute for Astronomy, the University of Hawaii, the Pan-STARRS Project Office, the Max-Planck Society and its participating institutes, the Max Planck Institute for Astronomy, Heidelberg and the Max Planck Institute for Extraterrestrial Physics, Garching, The Johns Hopkins University, Durham University, the University of Edinburgh, Queen’s University Belfast, the Harvard-Smithsonian Center for Astrophysics, the Las Cumbres Observatory Global Telescope Network Incorporated, the National Central University of Taiwan, the Space Telescope Science Institute, the National Aeronautics and Space Administration under Grant No. NNX08AR22G issued through the Planetary Science Division of the NASA Science Mission Directorate, the National Science Foundation under Grant No. AST-1238877, the University of Maryland, and Eotvos Lorand University (ELTE) and the Los Alamos National Laboratory. Based [in part] on data collected at the Subaru Telescope and retrieved from the HSC data archive system, which is operated by Subaru Telescope and Astronomy Data Center at National Astronomical Observatory of Japan.
%--
This work was partly done using GNU Astronomy Utilities (Gnuastro, ascl.net/1801.009) version 0.23. Work on Gnuastro has been funded by the Japanese Ministry of Education, Culture, Sports, Science, and Technology (MEXT) scholarship and its Grant-in-Aid for Scientific Research (21244012, 24253003), the European Research Council (ERC) advanced grant 339659-MUSICOS, the Spanish Ministry of Economy and Competitiveness (MINECO, grant number AYA2016-76219-P) and the NextGenerationEU grant through the Recovery and Resilience Facility project ICTS-MRR-2021-03-CEFCA. This work made use of \texttt{OverCite} \citep{Shariat2026}, an in-editor citation tool for \LaTeX.
%--
The Wendelstein 2.1m telescope project was funded by the Bavarian government and by the German Federal government through a common funding process. Part of the 2.1 m instrumentation including some of the upgrades for the infrastructure were funded by the Cluster of Excellence ‘Origin of the Universe’ of the German Science foundation DFG.
%--
The Hydrangea simulations were in part performed on the German federal maximum performance computer ‘HazelHen’ at the maximum performance computing centre Stuttgart (HLRS), under project GCS-HYDA / ID 44067, financed through the large-scale project ‘Hydrangea’ of the Gauss Center for Supercomputing. Further simulations were performed at the Max Planck Computing and Data Facility (MPCDF) in Garching, Germany.
%--Horizon
This work was granted access to the HPC resources of CINES under allocations 2013047012, 2014047012, and 2015047012 made by GENCI. This work has made use of the Infinity cluster, hosted by the Institut d’Astrophysique de Paris.The authors declare that they have no conflicts of interest. We thank the Hydrangea and Horizon collaborations for making their mock imaging products available for this work.

%%%%%%%%%%%%%%%%%%%%%%%%%%%%%%%%%%%%%%%%%%%%%%%%%%
\section*{Data Availability}

The datasets used in this work are publicly available. Hydrangea simulation data can be accessed at \href{https://ftp.strw.leidenuniv.nl/bahe/Hydrangea}{https://ftp.strw.leidenuniv.nl/bahe/Hydrangea},
HSC-SSP data at \href{https://hsc-release.mtk.nao.ac.jp/doc/}{https://hsc-release.mtk.nao.ac.jp/doc/}, and the K20 photometric profiles at \href{https://homepages.usm.uni-muenchen.de/mkluge/photometry.html}{https://homepages.usm.uni-muenchen.de/mkluge/photometry.html}.



\bibliographystyle{mnras}
\bibliography{example} % if your bibtex file is called example.bib

%\begin{thebibliography}{99}
%\end{thebibliography}

%%%%%%%%%%%%%%%%%%%%%%%%%%%%%%%%%%%%%%%%%%%%%%%%%%

%%%%%%%%%%%%%%%%% APPENDICES %%%%%%%%%%%%%%%%%%%%%

\appendix

%%%%%%%%%%%%%%%%%%%%%%%%%%%%%%%%%%%%%%%%%%%%%%%%%
%%%%%%%%%%%%%%%%%%%%%%%%%%%%%%%%%%%%%%%%%%%%%%%%%%

\section{Resolution tests}
\label{app:resolution}

To quantify the impact of resolution, as defined by pixel scale and spatial binning, we conducted a series of controlled tests. Below, we present the results for the K20 and Horizon-AGN samples, which we use as representative cases for observational and simulated data, respectively. We consider three  scenarios: (i) varying the spatial binning of the profiles, (ii) placing the same snapshot at higher redshift while keeping the same spatial sampling, and (iii) applying Gaussian smoothing.

\begin{figure}
  \centering
  \includegraphics[width=1\linewidth]{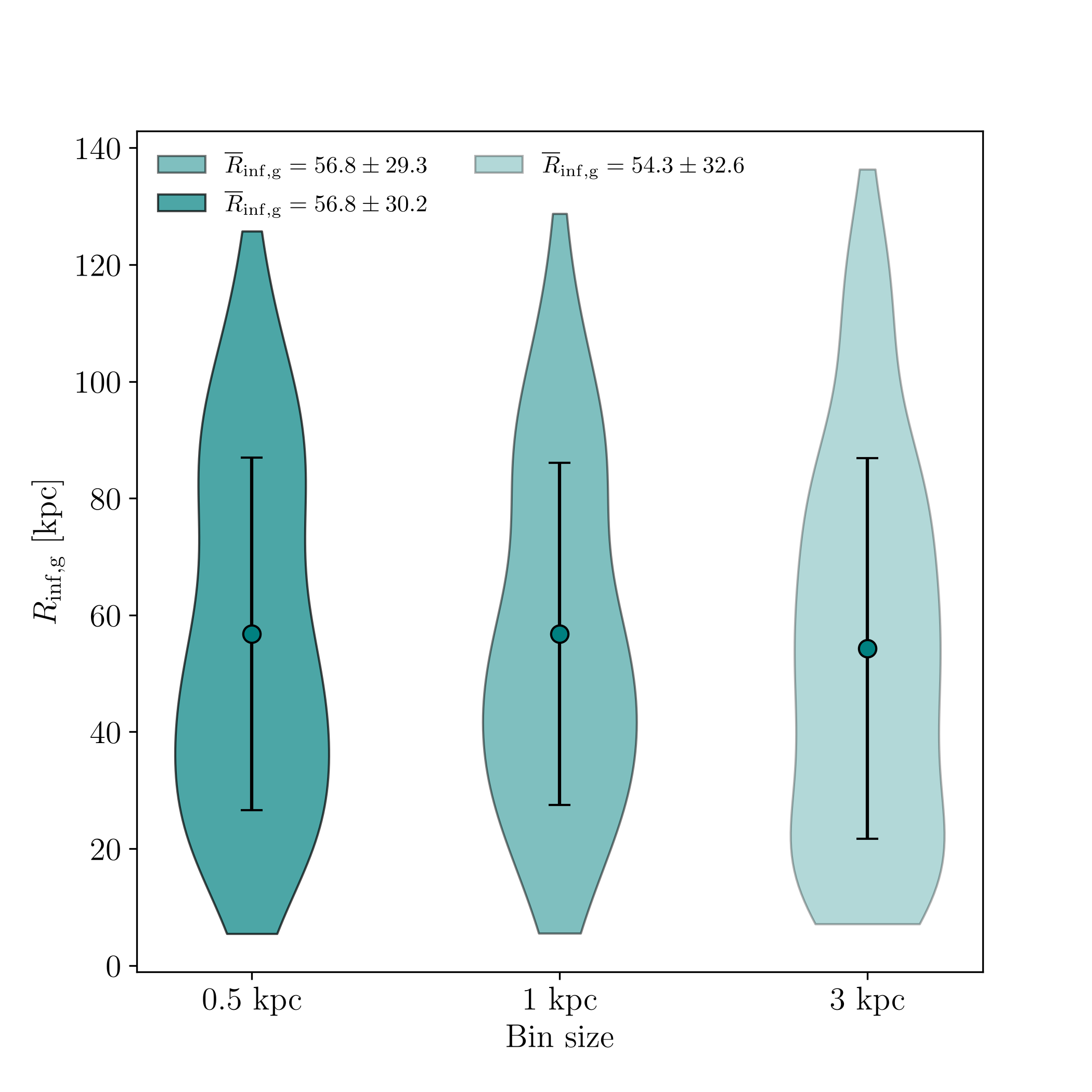}
  \caption{Testing different bin sizes when determining the inflection radius in K20 observed data. The violin plots show the inflection radius for each cluster in the K20 sample for three different bin sizes (0.5 kpc, 1kpc and 3 kpc). The mean values and their respective uncertainties are also shown.}
  \label{fig:diff_resolu_kluge}
\end{figure}

\textit{Spatial binning}: For the \cite{kluge2020structure} dataset, we first interpolate the surface brightness, PA, and $\epsilon$ profiles of each cluster. We then resample these profiles using bin sizes of $0.5$, $1$, and $3\,\mathrm{kpc}$. We adopt a resolution of $0.5\,\mathrm{kpc}$ per pixel as this approximates the highest effective resolution achievable in realistic observations, accounting for both instrumental effects and data processing. For comparison, LSST is expected to provide a physical resolution of $\sim$ 0.32 kpc/pix at $z = 0.1$ and $\sim$ 0.57 kpc/pix at $z=0.2$. Although higher resolution data (e.g.\ \textit{Euclid}) may allow for higher sampling, they could also introduce additional small-scale fluctuations (due to the presence of unmasked faint point-like sources), potentially resulting in the identification of multiple spurious inflection features.
At a redshift of $z = 0.4$ (the maximum redshift considered with HSC data), a pixel scale of $0.2''/\mathrm{pix}$ corresponds to $\sim$ $1\,\mathrm{kpc}$/pix. We then adopt $3\,\mathrm{kpc}$ as representative of observations of clusters at $z=0.4$ with SDSS-like data quality.
As displayed in Fig.~\ref{fig:diff_resolu_kluge}, the resulting inflection radii for the K20 dataset vary from 54.3 $\pm$ 32.6 kpc to 56.8 $\pm$ 30.2 kpc, demonstrating that the measurement is robust against variations in resolution within the uncertainties.

\begin{figure}
  \centering
  \includegraphics[width=\linewidth]{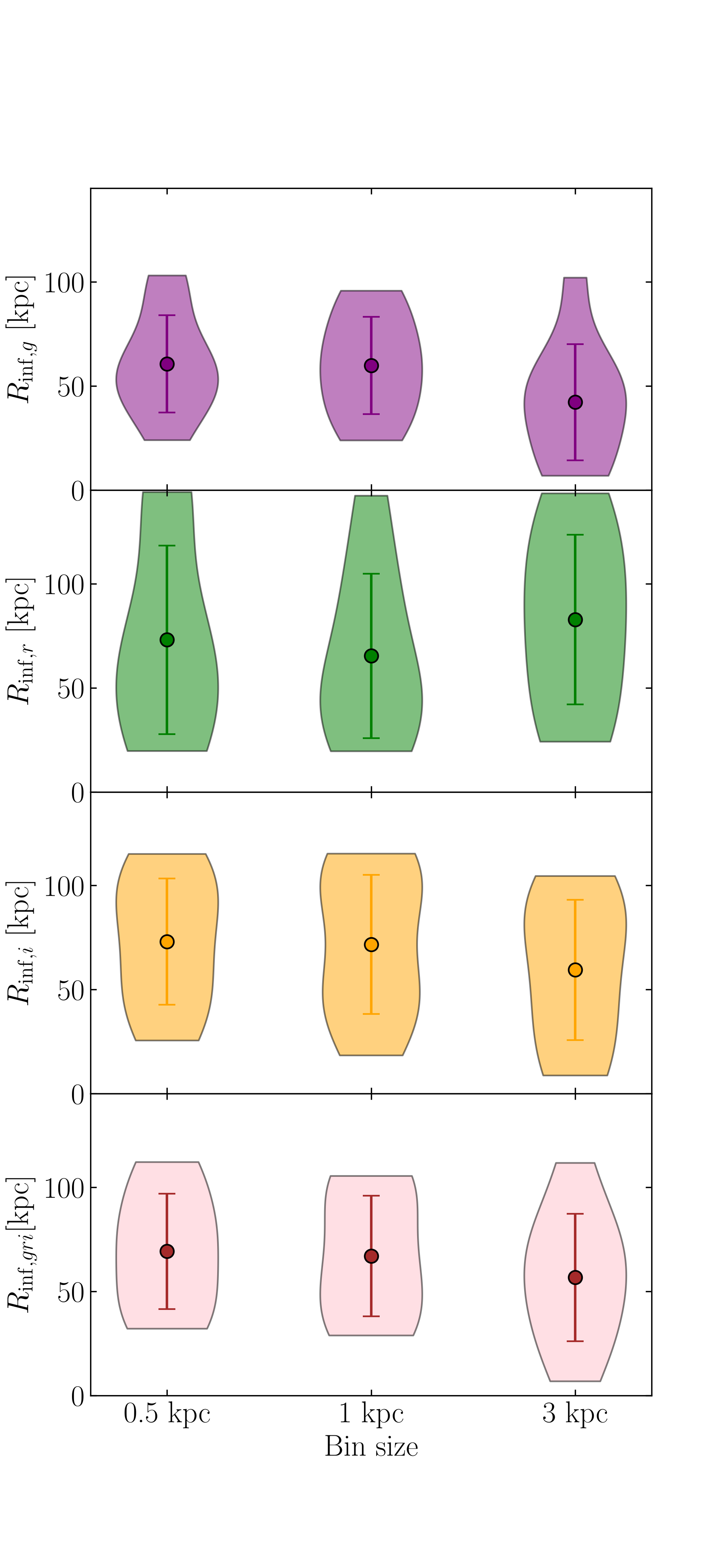}
  \caption{Variation in the results depending on the spatial binning of the surface brightness, PA and $\epsilon$ profiles for Horizon-AGN simulations. The distribution of the transition radii is shown for each cluster ID in the $g$, $r$, $i$ and $gri$ band, along with the corresponding mean and uncertainty for different spatial binning of the profiles: 0.5 kpc (left), 1 kpc (centre), 3 kpc (right).}
  \label{fig:diff-res}
\end{figure}

We further test the method on Horizon-AGN mock images, adopting bin sizes of $0.5\,\mathrm{kpc/pix}$, $1\,\mathrm{kpc/pix}$ and $3\,\mathrm{kpc/pix}$. The test is performed independently in the three photometric bands.
Figure~\ref{fig:diff-res} shows the inferred transition regions for each cluster at different resolutions, along with the corresponding distributions, for the three photoetric bands and $gri$. The mean values of the distributions are consistent within the uncertainties across all bands, indicating good agreement. We note, that the transition in the $g$ band is systematically recovered at smaller radii.
Overall, the inferred inflection radii remain stable across the range of tested resolutions. This behaviour is consistent with the expectation that the identified feature is not strongly influenced by spatial sampling.

\begin{figure}
  \centering
  \includegraphics[width=\linewidth]{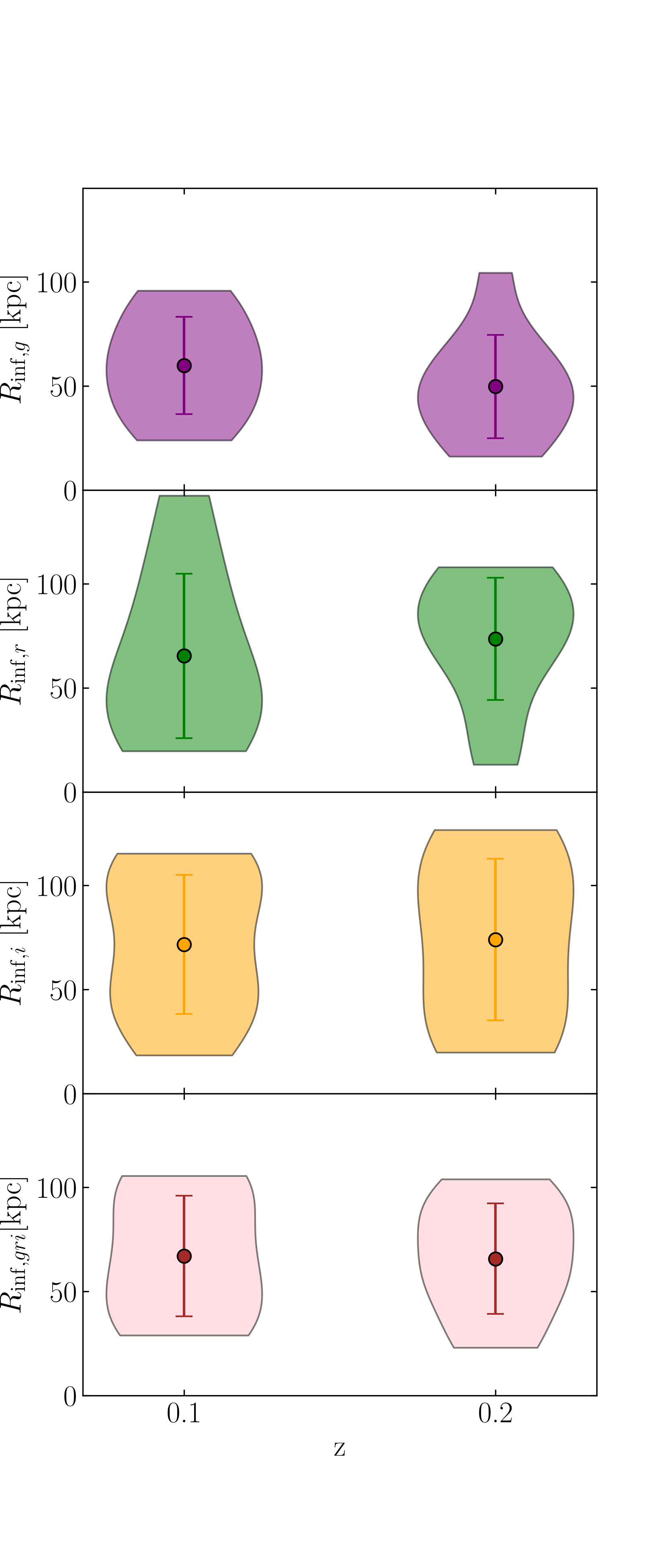}
  \caption{Distribution of the transition radii inferred from the inflection analysis for individual Horizon-AGN clusters in the $g$,$r$, $i$ and $gri$ bands at $z=0.1$ and $z=0.2$. Circles indicate the mean transition radius, obtained by averaging the measurements across the three bands. Error bars represent the uncertainty associated with the inferred inflection radius. }
  \label{fig:diffz1z2}
\end{figure}

\textit{Higher redshift}: An additional test made possible by the simulations involves placing the same snapshot at different redshift and verifying whether the inferred inflection radius is consistently recovered. Although this approach is conceptually similar to resampling at different bin sizes, it preserves the intrinsic structure of the mock images, which are generated directly at a given redshift ($z = 0.2$ in this case) with a fixed spatial resolution of 0.168$\arcsec$/pix. In this case, no physical shift in the transition radius is expected as the snapshot of the simulation is the same.
Figure~\ref{fig:diffz1z2} presents the results obtained by applying the methodology described in Sect.~\ref{sec:Methods} to Horizon-AGN mock clusters at $z=0.1$ and $z=0.2$ using a violin visualization in the $gri$ combined bands and $g$,$r$ and $i$ separately. The results are again consistent within the uncertainties, with the $g$ band tracing slightly smaller radii.

\begin{figure}
    \centering
    \includegraphics[width=1\linewidth]{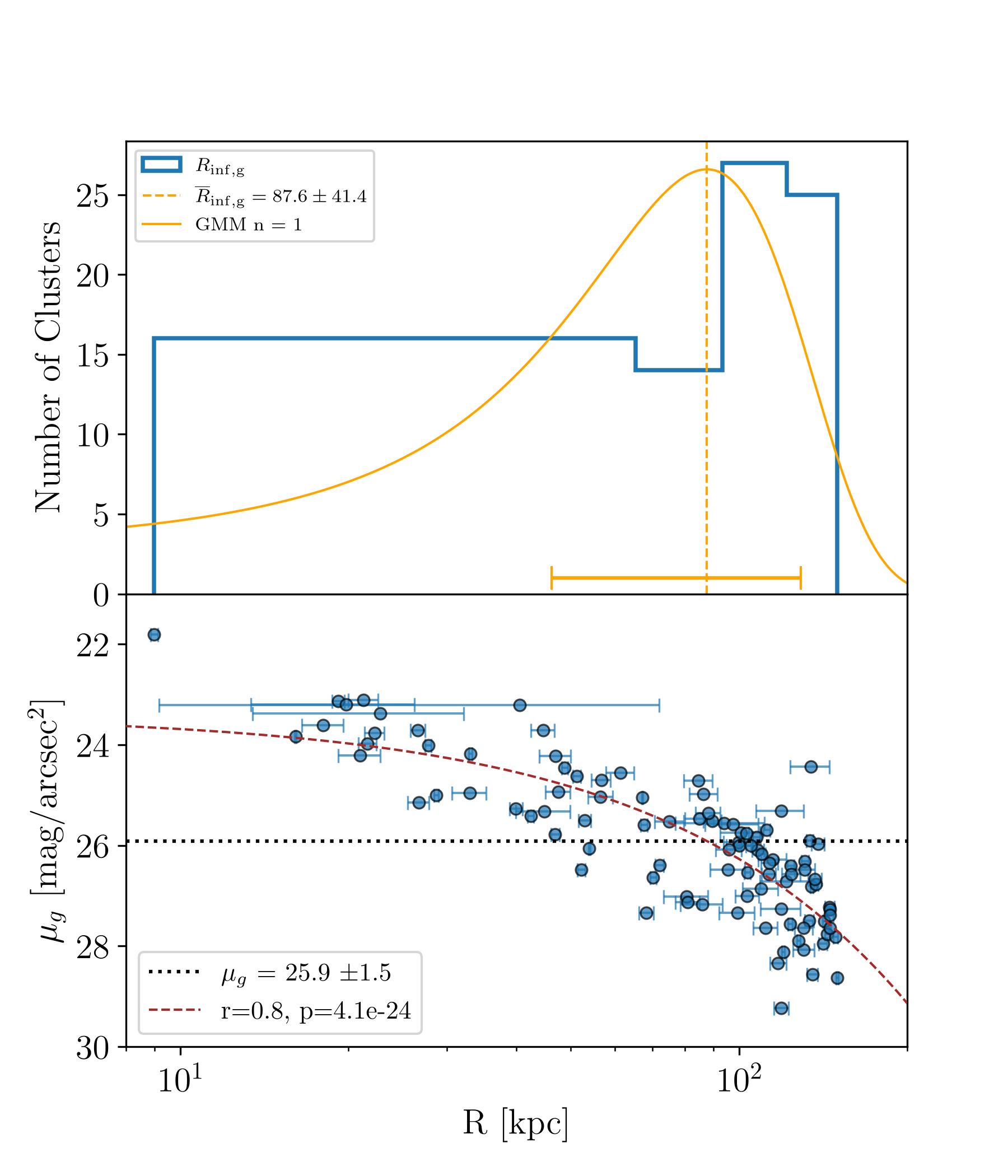}
    \caption{Similarly to Fig.~\ref{fig:hist-kluge}, distribution of the inflection radii detected for the 170 clusters in K20 using a Gaussian smoothing of 20 kpc. See text for details.}
    \label{fig:kluge20}
\end{figure}

\textit{Smoothing}: We also investigated the impact of smoothing on the profiles by applying the \texttt{gaussianfilter1d} function of Scipy \citep{nunez2017elegant} to the 170 clusters in K20.
We tested the effect of Gaussian smoothing using a kernel of $\sigma_g = 20$ kpc and present the resulting distribution in Fig.~\ref{fig:kluge20}. The new mean inflection radius is located at 87.6 $\pm$ 41.4 kpc with mean surface brightness of 25.9 $\pm$ 1.5 mag/arcsec$^{2}$.
As expected, smoothing reduces the amplitude of small-scale fluctuations in the second derivatives and decreases the number of candidate inflection points. 
Furthermore, applying Gaussian smoothing systematically shifts the location of the inflection point towards larger radii, as small-scale variations in the profiles are suppressed. For this reason and since our goal is to measure the transition radius directly from the observed profiles, we did not apply smoothing to the data in the main analysis, as this would have altered the intrinsic noise properties of the data. 

%%%%%%%%%%%%%%%%%%%%%%%%%%%%%%%%%%%%%%%%%%%%%%%%%%%%%%%
\section{Effect of the PSF}
\label{app:effectofpsf}

A critical challenge in deep galaxy photometry is the influence of the PSF, which redistributes light from the bright central regions of galaxies into their faint outskirts, potentially biasing the measurement of diffuse components. In the context of ICL studies, PSF contamination arises both from the extended wings of bright foreground stars and from the BCG, introducing artificial gradients and mimicking low surface brightness emission.
A range of approaches has been developed to mitigate these effects, including the construction of extended and custom PSF models to subtract scattered light on large spatial scales (e.g., \citealt{Gonzalez2005, kluge2020structure, montes2021buildup, martinez2023galaxy, bellhouse2025euclid, garate2025correcting}). In parallel, corrections for PSF-induced broadening of galaxy light have been implemented through direct subtraction of reconstructed scattered light, PSF-deconvolved 2D modeling of BCGs, and the use of high-resolution data to constrain the inner regions (e.g., \citealt{kluge2020structure, martinez2023galaxy}). PSF effects primarily impact the inner $10 - 20$ kpc, while becoming progressively negligible beyond $\sim 100$ kpc, with only minor influence on surface brightness profiles at large radii (\citealt{zhang2019dark, zhang2024dark, zhang2025splashback, bellhouse2025euclid}). Sensitivity analyses also indicate that inner structural parameters, such as the BCG effective radius, are most affected by PSF uncertainties (\citealt{Gonzalez2005}). To wrap up all these considerations, \cite{Golini2025} highlight the fundamental “PSF–stellar halo degeneracy,” emphasizing the difficulty in disentangling faint halos from PSF-scattered light. While the PSF alters profile slopes and outer brightness levels, it does not shift the location of the galaxy edge (therefore the structural parameters in BCG+ICL systems) and its impact depends on wavelength and instrumental characteristics.
Importantly, the present work focuses on radii beyond $\sim 20$ kpc, where PSF-induced flattening is most significant (\citealt{chen2022sphere, zhang2024dark}). This ensures that our measurements of the ICL and transition radius are less affected by PSF-related systematics over the radial range considered here. To further quantify these effects, we generate a suite of synthetic galaxy models to isolate the impact of PSF convolution on surface brightness profiles. The models, constructed with IMFIT \citep{erwin2015imfit}, an image-fitting tool designed for modeling galaxy surface brightness profiles, consist of Sérsic components (representing BCG+ICL -like profiles) with varying indices. For each configuration, images are created both with and without PSF convolution, and their radial profiles are extracted using ellipse fitting. By comparing these profiles, we directly assess how PSF smearing redistributes light as a function of Sérsic index and spatial sampling. We then compute the inflection radius following the procedure adopted in the main text and find that the inferred transition radius is robust against PSF effects. 
For Sérsic models with indices ranging from 1 to 10, the ratio between the transition radius measured before and after PSF convolution varies between 0.990 and 1.002, corresponding to a maximum change of less than 0.1$\%$.
Consistent results are obtained when an additional exponential component is included (Sérsic+Exponential), with a variation that decreases further, remaining below 0.01$\%$.

\section{Table of results}
\label{app:table}

In this appendix, we provide Table~\ref{Table:results}, which lists all inflection radii identified in this study for the K20, HSC, Horizon-AGN, and Hydrangea datasets across all available photometric bands.

\begin{table*}
  \centering
  \begin{tabular}{|l|c|c|c|c|c|}
   \toprule
   Dataset & Band & $R_{\rm inf}$ & $\mu(R_{\rm inf})$ & bin size&Note \\
     & & [kpc] & [mag/arcsec$^2$] & [kpc] &\\
  \midrule
  K20  & $g$ & 55.1 $\pm$ 29.0                  & 24.5 $\pm$ 1.4                        & 1     & 1S    \\
       & $g$ & 60.8 $\pm$ 29.7                  & 24.6 $\pm$ 1.0                        & 1     & 2S    \\
       & $g$ & $\meanKluge \pm \sigmameanKluge$ & $\mumeanKluge \pm \sigmamumeanKluge$  & 1     & 1S+2S \\
       & $g$ & 56.8 $\pm$ 30.2                  & 24.4 $\pm$ 1.2                        & 0.5   & 1S+2S \\
       & $g$ & 54.3 $\pm$ 32.6                  & 24.4 $\pm$ 1.3                        & 3     & 1S+2S \\
       & $g$ & 87.6 $\pm$ 41.4                  & 25.9 $\pm$ 1.5                        & 1     & 1S+2S, $\sigma_g$ = 20 \\
  \midrule
   HSC   & $g$   & $\meanhscg    \pm \sigmameanhscg$     & $\mumeanhscg \pm \sigmamumeanhscg$ & 1 & \\
         & $r$   & $\meanhscr    \pm \sigmameanhscr$     & $\mumeanhscr \pm \sigmamumeanhscr$ & 1 & \\
         & $i$   & $\meanhsci    \pm \sigmameanhsci$     & $\mumeanhsci \pm \sigmamumeanhsci$ & 1 & \\
         & $gri$ & $\meanhscgri  \pm \sigmameanhscgri$   &                                    & 1 & \\
  \midrule
  Horizon-AGN  & $g$ & $\meanhorizong \pm \sigmameanhorizong$ & $\mumeanhorizong \pm \sigmamumeanhorizong$ & 1   & z = 0.1  \\
               & $g$ & 49.7 $\pm$ 24.8                        & 24.2 $\pm$ 0.9                             & 1   & z = 0.2  \\
               & $g$ & 60.6 $\pm$ 23.3                        & 24.0 $\pm$ 1.1                             & 0.5 & z = 0.1  \\
               & $g$ & 42.2 $\pm$ 27.9                        & 23.1 $\pm$ 1.5                             & 3   & z = 0.1  \\
  
               & $r$ & $\meanhorizonr \pm \sigmameanhorizonr$ & $\mumeanhorizonr \pm \sigmamumeanhorizonr$ & 1 & z = 0.1  \\
               & $r$ & 73.5 $\pm$ 29.4                        & 23.9 $\pm$ 1.0                             & 1 & z = 0.2\\
               & $r$ & 73.1 $\pm$ 45.2                        & 23.3 $\pm$ 1.4                             & 0.5 & z = 0.1  \\
               & $r$ & 82.8 $\pm$ 40.7                        & 23.5 $\pm$ 1.2                             & 3 & z = 0.1  \\
              
               & $i$ & $\meanhorizoni \pm \sigmameanhorizoni$ & $\mumeanhorizoni \pm \sigmamumeanhorizoni$  & 1 & z = 0.1  \\
               & $i$ & 74.0 $\pm$ 38.8                         & 23.4 $\pm$ 1.4                             & 1 & z = 0.2\\
               & $i$ & 73.1 $\pm$ 30.4                         & 23.0 $\pm$ 1.3                             & 0.5 & z = 0.1  \\
               & $i$ & 59.5 $\pm$ 33.7                         & 23.4 $\pm$ 1.5                             & 3 &  z = 0.1  \\
               
               & $gri$ &  $\meanhorizongri \pm \sigmameanhorizongri$  & &1 & z = 0.1  \\
               & $gri$ &  65.7 $\pm$ 26.5                             & &1 & z = 0.2  \\
  \midrule
  Hydrangea    & $g$  & $\meanhydrangeag \pm \sigmameanhydrangeag$         & $\mumeanhydrangeag \pm \sigmamumeanhydrangeag$ & 1 &z = 0.1  \\
               & $r$  & $\meanhydrangear \pm \sigmameanhydrangear$         & $\mumeanhydrangear \pm \sigmamumeanhydrangear$ & 1 &z = 0.1  \\
               & $i$  & $\meanhydrangeai \pm \sigmameanhydrangeai$         & $\mumeanhydrangeai \pm \sigmamumeanhydrangeai$ & 1 &z = 0.1  \\
               & $gri$& $\sigmameanhydrangeagri \pm \sigmameanhydrangeagri$&                                                & 1 &z = 0.1  \\
  \bottomrule

  \end{tabular}
  \label{Table:results}
  \caption{Summary of the results presented in this work for different datasets in different configurations.}
\end{table*}

\section{1S and 2S histograms of the K20 sample}
\label{app:klugehist12}

\begin{figure}
  \centering
  \includegraphics[width=1\linewidth]{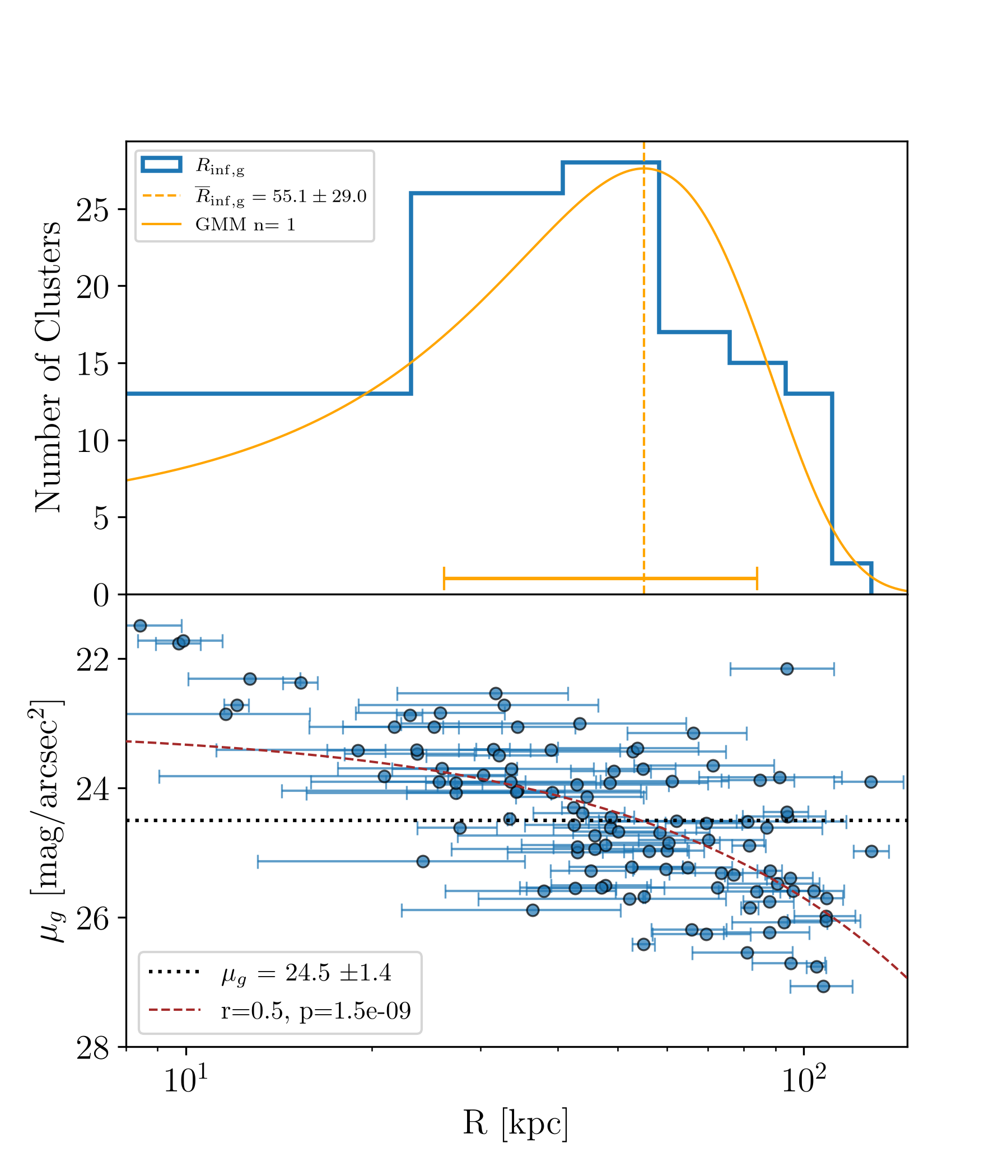}
  \caption{Similarly to Fig.~\ref{fig:hist-kluge}, distribution of the characteristic inflection radii for the single Sérsic cluster sample in K20. See main text for details.}
  \label{fig:kluge1sersic}
\end{figure}

\begin{figure}
  \centering
  \includegraphics[width=1\linewidth]{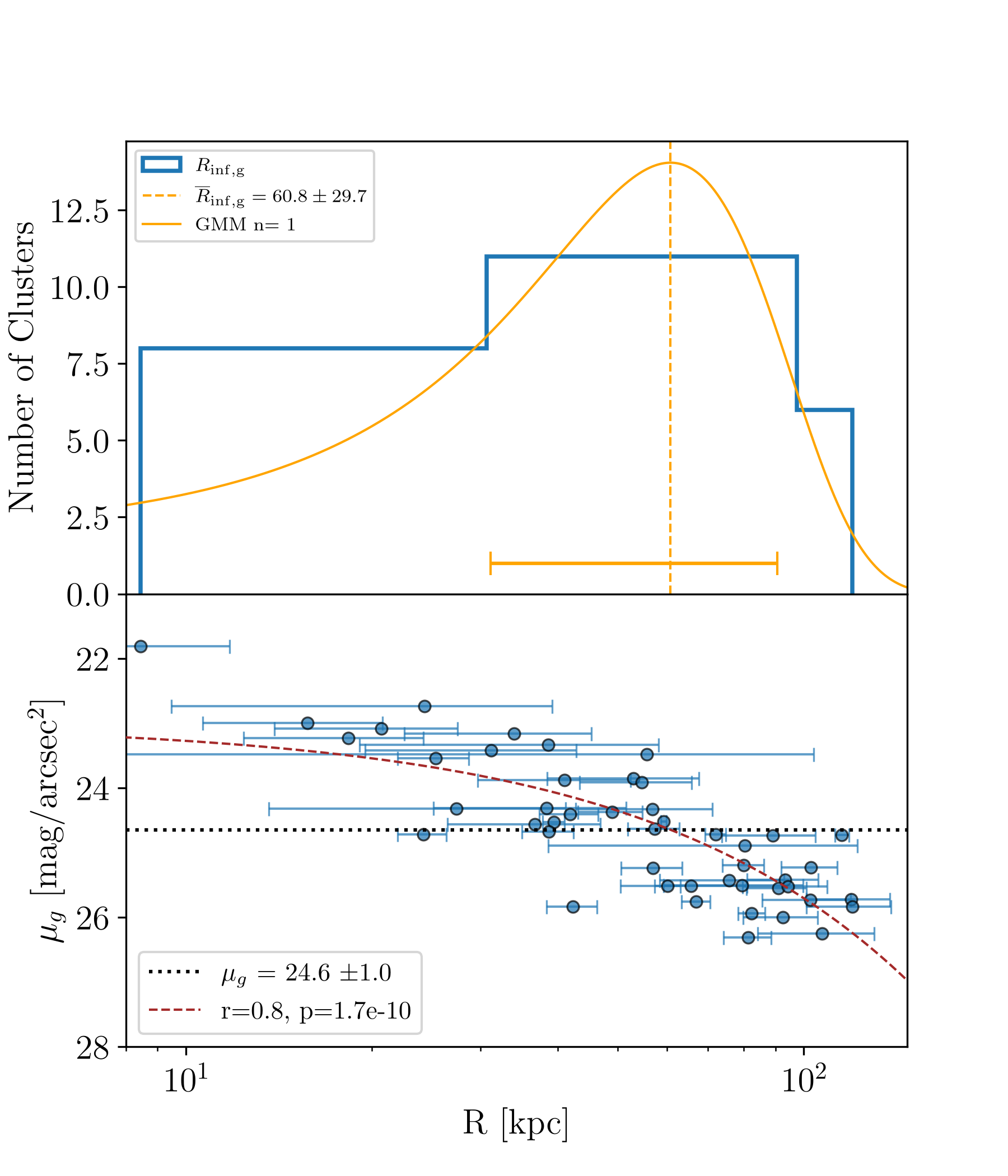}
  \caption{Similarly to Fig.~\ref{fig:hist-kluge}, distribution of the characteristic inflection radii for the double Sérsic cluster sample in K20. See main text for details.}
  \label{fig:kluge2sersic}
\end{figure}

In this Appendix, in Fig.~\ref{fig:kluge1sersic} and Fig.~\ref{fig:kluge2sersic} we report the histograms of the distributions of the inflection radii in K20 divided following the description in \cite{kluge2020structure} for a single and double Sérsic components modeling respectively. 
In both Fig.~\ref{fig:kluge1sersic} and Fig.~\ref{fig:kluge2sersic} (top panels), the vertical dashed line and error bar indicate the mean and, the corresponding standard deviation is shown as a horizontal bar. Overlaid is a one component GMM fit, scaled to the histogram counts. In the bottom panel of both figures is shown, for each cluster, the surface brightness as a function of the transition radius, with corresponding uncertainties. The dashed line represents a linear Pearson fit to the data. The mean of the surface brightness distribution is displayed as a horizontal dotted line.

\section{Bands histograms}
\label{app:hist-individual bands}

\begin{figure*}
  \centering
  \includegraphics[width=1\linewidth]{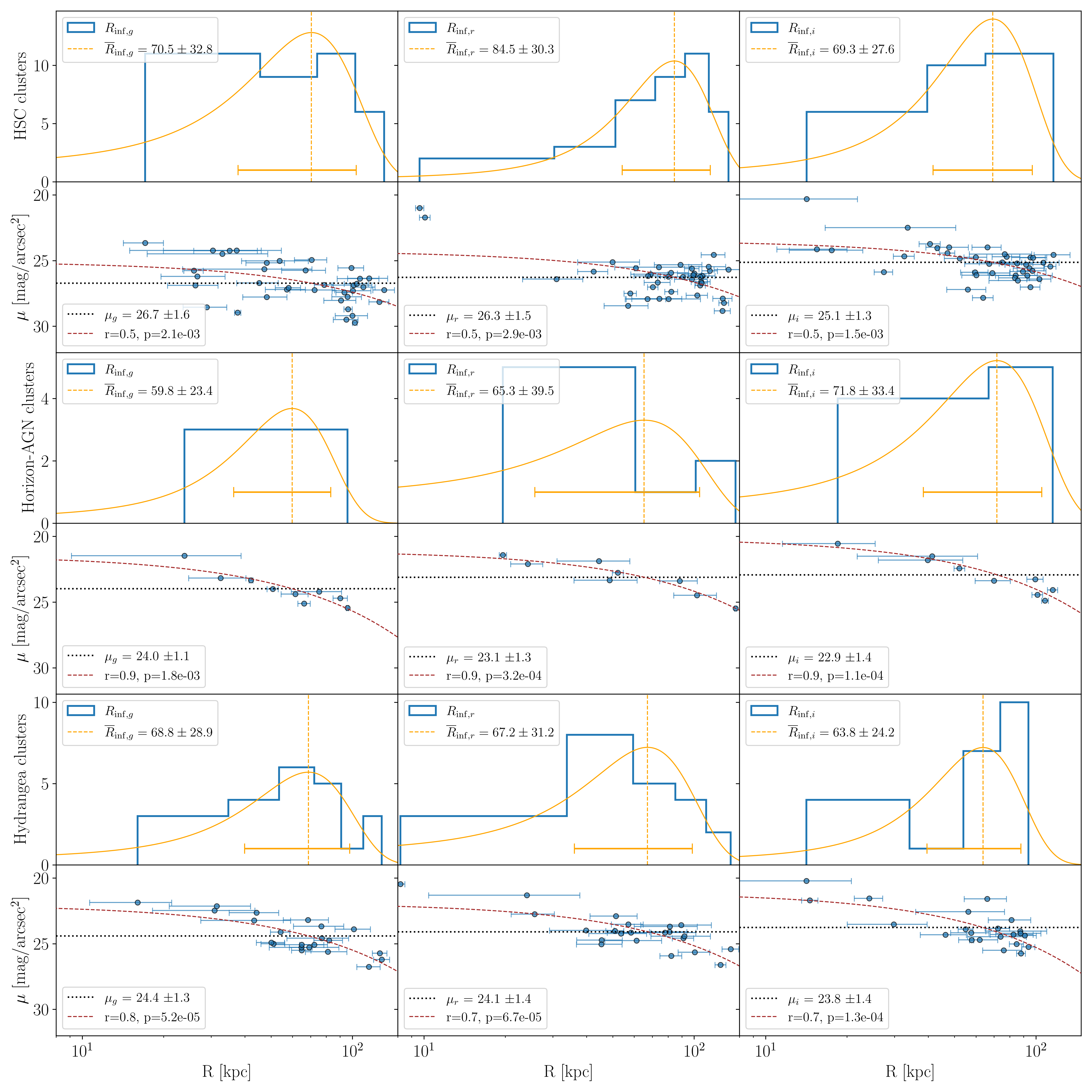}
  \caption{Top row: Distributions of the characteristic radius, $R_{\mathrm{inf},f}$, in the $g$-, $r$-, and $i$-bands for the HSC cluster sample. Dashed lines and error bars denote the $3\sigma$-clipped mean and standard deviation of each distribution, respectively. Solid curves show single-component Gaussian mixture model fits, scaled to the histogram counts. Second row: Surface brightness at the transition radius as a function of $R_{\mathrm{inf},f}$ for individual HSC clusters. Dashed lines indicate linear fits to the data, while horizontal dotted lines mark the mean surface brightness of each distribution. Legends report the Pearson correlation coefficient (r) and corresponding p value for each band. The third and fourth rows are analogous to the first and second rows, but for the Horizon-AGN cluster sample, while the fifth and sixth rows show the corresponding distributions and relations for the Hydrangea mock cluster sample.}
  \label{fig:hist_ZeroCrossings_all}
\end{figure*}

Fig.~\ref{fig:hist_ZeroCrossings_all} summarises the distribution and photometric properties at the inferred inflection radius for three independent datasets (split in couple of rows) and three photometric bands (columns: $g$, $r$, and $i$). The upper panels of each dataset show the distribution of transition radii, $R_{\mathrm{inf}}$, while the lower panels display the corresponding surface brightness, $\mu$, measured at $R_{\mathrm{inf}}$ as a function of radius.

In the histogram panels, the histograms show the distribution of transition radius values across the cluster samples, while the solid curves correspond to Gaussian fits. The vertical dashed lines on top of the histograms indicate the mean transition radius, and the horizontal error bars provide a visual representation of the associated dispersion. Across all datasets, the distributions are broadly consistent with a unimodal shape. The scatter varies between samples and bands.

The scatter plots below each histogram show the surface brightness at the transition radius for each individual cluster. Error bars reflect measurement uncertainties, while the dashed lines indicate linear fits to the data. The associated Pearson correlation coefficients are reported in the legend. The dotted horizontal lines mark the mean surface brightness values for each band. 

Comparing the rows highlights the differences between the observational (HSC) and simulated samples (Horizon-AGN and Hydrangea). While the inferred transition radii are broadly consistent across all datasets, the simulations exhibit brighter mean surface brightness values. 

%%%%%%%%%%%%%%%%%%%%%%%%%%%%%%%%%%%%%%%%%%%%%%%%%%%%%%%%%%%%

% Don't change these lines
\bsp	% typesetting comment
\label{lastpage}
\end{document}